\documentclass[aps,prb,floatfix,showpacs,amssymb,amsmath]{revtex4}
\usepackage{graphicx}% Include figure files
\usepackage{epsfig}
\usepackage{dcolumn}% Align table columns on decimal point
\usepackage{bm}% bold math

\newcommand{\ket}[1]{\left| #1 \right\rangle}
\newcommand{\bra}[1]{\left\langle #1 \right|}

\begin{document}

\title{Phonon-assisted decoherence  in  coupled  quantum dots}  
\author{A. Thilagam and M. A. Lohe}
\affiliation{Department of Physics, \\ 
 The University of Adelaide, Australia
 5005}
\date{\today }

\begin{abstract}
We analyse various phonon-assisted mechanisms which  contribute to the
decoherence  of excitonic qubits in  quantum dot systems coupled by the 
F\"orster-type transfer process. We show the significant loss of coherence
 accompanied by dissipation due to  charge carrier oscillations between qubit 
states by using a model of one-phonon assisted F\"orster-type transfer process. 
We  obtain explicit expressions for the relaxation and
dephasing times for excitonic qubits interacting with 
acoustic phonons  via both deformation potential and piezoelectric coupling.
 We compare the decoherence times of the GaAs/AlGaAs 
material system with half times of
 concurrence decay in a 2$\otimes$2  bipartite mixed
excitonic qubit system. We extend calculations to determine 
the influence of phonon mediated interactions on the non-unitary
evolution  of Berry phase in quantum dot systems.

\end{abstract}
%73.20.Mf  Excitons in quantum dots: electronic structure and optical properties
%78.67.Hc Quantum dots: optical properties
%03.67.Lx Quantum computation
%89.70.+c Information science
% 85.35.Gv   Single electron devices

\maketitle

\section{\label{intro}Introduction}

The use of excitons (electron-hole correlated  states) in 
quantum dots has generated great interest as  suitable 
devices for the solid-state implementation of  quantum logic gates 
\cite{Loss,Biolatti,Science,Krenner}. The significant  advancement in
creating and probing excitonic states in quantum dots \cite{mull,zre} and
newly developed techniques  which allow quantum  coupling
 between excitonic states to be continuously varied by  
external electric fields \cite{Krenner} has provided rapid progress in the
 field of solid-state quantum computation. 
In the case of two coupled quantum dot systems,
exciton-exciton dipole interactions  give
 rise to diagonal terms which allows  quantum logic to be 
performed via ultra-fast laser pulses on time scales 
less than calculated decoherence times\cite{Biolatti}.
Related studies \cite{love,Nazir,Briggs} have suggested
the importance of the F\"orster energy transfer process 
in producing  entangled excitonic states and using
external electric fields to achieve critical logic gate actions.
Dipole-dipole interaction was first 
proposed by F\"orster\cite{fos} and further extended by 
Dexter\cite{dex} as a  mechanism of energy  transfer in which 
excitonic  wavefunctions localized at different  quantum dots
do not overlap.

Other than the actual generation of entangled excitonic qubit states,
the successful implementation of quantum logic systems 
depends critically on the decoherence properties of the qubits.
Decoherence  due to enviromental factors such as phonons and 
 impurities is inevitable and is regarded as a  major drawback 
 in solid-state devices. Several strategies  such as 
decoherence-free subspaces \cite{df1,df2},  optimal control
techniques \cite{hon} and immunization processes \cite{xue}
have been proposed to counter the detrimental effects
of decoherence. Geometric quantum computation 
has gained increased status  as a robust fault-tolerant scheme 
in recent years \cite{zan1,zan2}. 
The geometric phase component (Berry phase) of the 
phase shift in quantum states  relies solely on the global geometry of the
 path executed during cyclical evolution 
 and hence is manifestly gauge invariant and consequently
 insensitive to stochastic operation errors \cite{zan3}. 
It is not immediately clear however whether  these  proposals invoke other  
routes  which facilitate decoherence and if so the role
played by various system parameters in minimizing errors during logic gate operations.
Hence there is a need to scrutinize the various intrinsic processes
and in particular phonon induced types in order to achieve
  fault tolerant quantum switches.  

The interaction of charge carriers with acoustic phonons arises mainly from
deformation-potential and piezoelectric coupling  in quantum dots \cite{Takag}. 
Theoretical estimates by Fedichkin at al \cite{Fed} show that 
the error rate due to acoustic phonons may be a major factor 
limiting qubit performance, in agreement with experimental
results \cite{hay} of a qubit system implemented using a 
double-dot GaAs/AlGaAs system with a two dimensional electron gas. 
Electron-phonon interaction has been shown be significant
in GaAs/InAs quantum dots compared to the bulk system of the same
material leading to the formation of polarons \cite{jak1}.
In another work, Jacak and coworkers have studied the
significance of relaxation of electrons surrounded by 
a cloud of phonons due to the anharmonic interaction of optical phonons
with acoustic phonons in GaAs/InAs quantum dot systems \cite{jak2}.
A recent study \cite{Mach} has considered the possibility of
localization\cite{zurek} in which an electron state 
delocalized over two quantum dots is converted into a 
mixture of two classical-like localized states  and 
consequently loses its coherence to the surrounding reservoir of phonons.

The interaction of qubits with phonons is generally modelled 
as a first-order process taking into account only  
the coupling of carrier  states localized in
 quantum dots \cite{old} in contact with a phonon bath.
The  interaction of charge carriers 
with phonons as the qubit flips back and forth between the
quantum dots has not been well investigated 
in earlier studies related to phonon mediated
decoherence in semiconductor systems \cite{Fed,cli,wu}. 
The loss of coherence due to charge carrier movements
 during oscillations between the different
 qubit states is unavoidable, and therefore we aim
to closely examine this critical mode of decoherence in this work.
We model this route of  degradation in the  
purity and entanglement of qubit states to proceed
via a one-phonon assisted oscillation of qubit states 
leading to relaxation as well as dephasing of the qubit states.
In order to obtain explicit results, we
consider the specific case of excitonic qubits in F\"orster 
coupled quantum dots \cite{Nazir}.  An earlier work \cite{Briggs}
has examined the coupling of the F\"orster coupled quantum 
dot system to the external environment consisting
 of a single radiation mode and the associated
spontaneous emission and decay processes. We are unaware of  other
detailed studies of phonon mediated interactions involving excitonic 
 qubit states in  F\"orster coupled  quantum dots.

It is not immediately clear  whether decoherence phenomenon is dominated
by relaxation or pure  dephasing process and the extend to which the outcome is
affected by the nature of qubit-phonon coupling  during a  cycle of logic 
gate operation. The subtle difference between
relaxation and pure dephasing of entangled systems has to be emphasized.
While relaxation brings about shifts in the population of qubits states
with one state favored over another, pure dephasing process results
in changes in the energy difference between  qubit states with the 
population of qubit states remaining intact.
 While both processes lead to decoherence of the entangled 
system, we aim to examine the  various parameters affecting the two inherent 
sources of dissipation  in this work. Our  approach undertaken here
can be extended to other designs of coupled dot qubit systems such
as the electrically controlled charge qubit in the double quantum dot
system or the double impurity semiconductor systems \cite{clark}.
 
It is well known that a quantum system exposed to an environment 
is best described by the reduced density matrix for which 
approximation techniques \cite{priv} have been developed to describe 
the quantum state evolution at times determined by the environmental noise.
Non-Markovian stochastic Schrodinger equations \cite{Vega}, 
Lindblad \cite{potz} and Redfield master equations \cite{esp} 
have  been independently used  
to investigate the dynamics of  open quantum  system in contact  
with  noisy environments. In a recent work, Braun \cite{braun}
showed  that two qubits  not interacting directly but 
which are  in contact with a common environment
can become entangled and the entanglement may last for a long time depending 
the strength of coupling and environmental conditions. These results
are similar to those obtained by Ficek et al \cite{fic} who  studied a
system of  two suitably prepared atoms immersed in a vacuum field. The term
"sudden death" has been coined in an earlier
work \cite{Yu} to describe the process in which two initially entangled
atoms which become noninteracting later become completely disentangled in 
relatively short times. The presence of such features have not been thoroughly
investigated in excitonic qubit systems.

In the second part of this work, we investigate  the
 dynamical behaviour of entanglement of excitons
 coupled to  two phonon baths associated with 
deformation-potential and piezoelectric coupling in the absence of
a direct channel of interaction, i.e the F\"orster interaction
is switched off by control of external electric fields.
We follow the approach of Tolkunov et al \cite{arvind} in which  the
degree of loss of coherence is determined by the decay of 
of the off-diagonal terms of the reduced density matrix.
We determine two important quantities, namely
the Berry phase and concurrence \cite{woott}
and show  the strong correlation between the 
Berry phase correction fraction which quantify departure from unitary
 evolution and half times of concurrence decay 
 in GaAs/AlGaAs quantum dots.

\section{\label{exci}Excitonic qubits in quantum dots}

We consider two  excitons in their ground states
 in adjacent coupled quantum dots  located
at ${\bf R_{\rm a}}$ and  ${\bf R_{\rm b}}$. We assume the quantum 
dots to be shaped in the form of either cuboid boxes or
quasi-two dimensional disks in which the vertical
 confinement energies of charge carriers 
are larger than their lateral confinement energies. 
 We label the localized excitonic states as $\ket{\bf R_{\rm a}}$ 
and  $\ket{\bf R_{\rm b}}$. Employing some of the notations in 
our earlier works \cite{thilaM}, we represent  $\ket{\bf R_{\rm a}}$ as
\begin{equation}
\label{eq:istate}
\ket{\bf R_{\rm a}} =  \frac {\nu_0}{L} \sum_{\bf r_e,r_h}
\; \Phi({\bf R_{\rm a},r_{e \parallel},r_{h \parallel}},
z_e,z_h)\;
 \; a_{c,{\bf r_e}}^{\dagger}
\;a_{v,{\bf r_h}}^{} \; \ket{\bf 0},
\end{equation}
where  $\nu_0$ is the volume of the unit cell, $L$ is the quantization
length  and $a_{c,{\bf r_e}}^{\dagger}$ ($a_{v,{\bf r_h}}^{}$) is
the creation (annihilation) operator of an  electron  
in  the conduction (valence) band, denoted by $c$ ($v$).
$\ket{\bf 0}$ in Eq. \ref{eq:istate} denotes the electronic state 
of the quantum dot in which all electronic ground states are occupied 
and all excited states are unoccupied. 
For simplicity, we ignore the role of spin effects and 
spin indices in Eq. \ref{eq:istate}. 
 The position vectors ${\bf r_e}$ and ${\bf r_h}$ 
are decomposed into components parallel and perpendicular 
to the lateral direction of the quantum dot as
${\bf r_e} = ({\bf r_{e \parallel}}, z_e)$ and ${\bf r_h}
 = ({\bf r_{h \parallel}},z_h)$.
The vertical confinement energies of charge carriers are
stronger than the lateral confinement energies, the exciton wavefunction
and therefore we factorise 
 $\Phi({\bf R_{\rm a},r_{e \parallel},r_{h \parallel}},z_e,z_h)$ as
\begin{equation}
\Phi({\bf R_{\rm a},r_{e \parallel},r_{h \parallel}},z_e,z_h) = 
\Psi({\bf R_{\rm a},r_{e \parallel},r_{h \parallel}})\; 
\varphi_e(z_e) \; \varphi_h(z_h),
\label{eq:exfunc1}
\end{equation}
where $\varphi_e(z_e)$ ($\varphi_h(z_h)$) is the envelope function of the
electron (hole) in the vertical direction of the quantum dot. 
The form of the in-plane exciton wavefunction 
$\Psi({\bf R_{\rm a},r_{e \parallel},r_{h \parallel}})$ at
${\bf R_{\rm a}}$ depends on the degree of confinement of the
electron-hole within the quantum dot.  We consider a strong 
individual charge carrier regime
in which the kinetic motions of the electron and holes
are quantized separately so that the resulting discrete energy levels are 
affected by the Coulomb interaction between the electron and hole 
only by a small amount of the order $\approx \frac{1}{L}$ where $L$ is
the quantum dot radius. 

$\Psi({\bf R_{\rm a},r_{e \parallel},r_{h \parallel}})$ is further factorized 
using a suitable change of coordinates 
\begin{eqnarray}
\nonumber
\label{eq:coords}
{\bf r_\parallel} & = &  \frac{1}{\rm l_r} ({\bf r_{e \parallel} -
 r_{h \parallel}})\\ \nonumber
{\bf R} & = &  \frac{1}{\rm l_r^2} ({\rm l_h^2} \; {\bf r_{e \parallel}} -
{\rm l_e^2} \; {\bf r_{h \parallel}})
\end{eqnarray}
where ${\rm l_r} = \sqrt{{\rm l_e}^2+{\rm l_h}^2}$. 
The effective lengths ${\rm l_e}$ and ${\rm l_h}$ are  related to the respective 
electron and hole effective masses and 
confining potential frequencies, $\omega_o^e$
and $\omega_o^h$ by ${\rm l_e} = \sqrt{\frac{\hbar}{m_e \omega_o^e}}$ 
and ${\rm l_h} = \sqrt{\frac{\hbar}{m_h \omega_o^h}}$. 
We obtain  $\Psi({\bf R_{\rm a},r_{e \parallel},r_{h \parallel}}) 
= \Xi({\bf R, R_{\rm a}}) \;
\psi({\bf r_\parallel})$ where the confining potential in the lateral 
direction are modelled using harmonic potentials 
\begin{eqnarray}
\label{eq:exciwavefnlat}
\psi({\bf r_\parallel}) &=&  \frac{1}{\sqrt{\pi} {\rm l_r}} 
\exp(- \frac{r_\parallel ^2}{2 {\rm l_r}^2}), \nonumber \\
\Xi({\bf R, R_{\rm a}}) &=&  \frac{1}{\sqrt{\pi}{\rm L_R}} 
\exp(- \frac{1}{2 {\rm L_R}^2}{\bf |R_\parallel -  R_{\rm a}|}^2)
\end{eqnarray}
where ${\rm L_R}^2 = {\rm l_e}^2  {\rm l_h}^2/{\rm l_r}^2$.
 The  length scales ${\rm l_e}$ and ${\rm l_h}$ associated with the
harmonic potential  are smaller than the effective Bohr radius of the exciton
due to the strong confinement of charge carriers in the quantum dot.

The Gaussian form of function $\Xi({\bf R, R_{\rm a}}) $ in
 Eq. \ref{eq:exciwavefnlat} yields an explicit expression
for its two-dimensional Fourier transform:
\begin{eqnarray}
\label{eq:fouriertransform}
\Xi({\bf R, R_{\rm a}}) &=&  \int \! d^2 {\bf q_\parallel} \;
\exp({\rm i} {\bf q_\parallel \cdot R}) \;  
\; \tilde{\Xi}({\bf R, R_{\rm a}}), \nonumber \\
\tilde{\Xi}({\bf R, R_{\rm a}}) &=&  \frac{{\rm L_R}}{2 \pi \sqrt{\pi}} 
\exp(- {\rm i} {\bf q_\parallel \cdot 
R_{\rm a}} - \frac{{\rm L_R}^2 q_\parallel^2}{2} )
\end{eqnarray}

The confining potential in the 
 vertical direction are modelled using harmonic potentials
 defined by the vertical confinement dimensions ${\rm l_{ze}}$ 
and ${\rm l_{zh}}$ respectively for electrons and holes
\begin{eqnarray}
\label{eq:exciwavefnvert}
\varphi_e(z_e) &=&  (\frac{1}{\sqrt{\pi} {\rm l_{ze}}})^{\frac{1}{2}}
\exp(- \frac{z_e^2 }{2 {\rm l_{ze}}^2}), \nonumber \\
\varphi_h(z_h) &=&  (\frac{1}{\sqrt{\pi} {\rm l_{zh}}})^{1/2}  
\exp(- \frac{z_h^2 }{2 {\rm l_{zh}}^2}),
\end{eqnarray}
The excitonic state $\ket{\bf R_{\rm b}}$ 
is  analogous  in form to  Eq. \ref{eq:istate}.

We code the excitonic qubits states using  
the relative position of the exciton via the  
basis set, ($\ket{\bf L}, \ket{\bf R}$) 
\begin{eqnarray}
\nonumber
\label{eq:qstates}
\ket{\bf L}  & = &  \ket{\bf R_{\rm a}} \otimes 
 \ket{\bf 0}_b \\
\ket{\bf R}  & = & \ket{\bf 0}_a \otimes \ket{\bf R_{\rm b}},
\end{eqnarray}
The states, $\ket{\bf 0}_a$ and $\ket{\bf 0}_b$, which
correspond to the  absence of excitons denote 
the respective ground states of the quantum dots at
$\ket{\bf R_{\rm a}}$ and  $\ket{\bf R_{\rm b}}$.
We simplify the approach by considering a two-level system
involving only the states $\ket{\bf L}$ and $\ket{\bf R}$
and work in the limit of a pure F\"orster  coupling. The
direct Coulomb interaction which causes the formation of the
 biexciton state $\ket{\bf R_{\rm a}} \ket{\bf R_{\rm b}}$
is neglected. We also exclude  the possibility of entangled states 
involving the vacuum state, $\ket{\bf 0}_a \ket{\bf 0}_b$. 
The two level excitonic qubit Hamiltonian is written as
\begin{equation}
\label{eq:Hamqubit}
\hat H_{\rm ex-qb}
= - \hbar (\frac{\Delta \Omega}{2}\, \sigma_{z} + F \, \sigma_{x}),
\end{equation}
where the Pauli matrices $\sigma_{z} = \ket{\bf L} \bra{\bf R}
+ \ket{\bf R} \bra{\bf L}$ and $\sigma_{z} = \ket{\bf L} \bra{\bf L}
- \ket{\bf R} \bra{\bf R}$. $\Delta \Omega =  \Omega_{\rm a} - \Omega_{\rm b}$ is the 
difference in  exciton creation energy between the 
quantum dot at ${\bf R_{\rm a}}$ and that at 
${\bf R_{\rm b}}$. $F$  denotes the interdot
F\"orster interaction amplitude responsible for  the
transfer of an exciton from one quantum dot to the other without
involving  a tunnelling process.  

The eigenstates  of the interacting  
qubit system in Eq. \ref{eq:Hamqubit} appear as
\begin{eqnarray}
\label{eq:statesEigen}
\ket{\chi_{\rm s}}  & = & \cos(\beta /2)
 \ket{\bf L} \; + \sin(\beta /2)\ket{\bf R}  \\ \nonumber
\ket{\chi_{\rm as}}  & = & \sin(\beta /2)
 \ket{\bf L} \; - \cos(\beta /2)\ket{\bf R},
\end{eqnarray}
 with  eigenenergies
\begin{eqnarray}
\label{eq:energyEigen}
E_{\rm as}  & = & \Omega_0 + \Omega_1 - \frac{\Delta \Omega}{2} - 
\sqrt{(\frac{\Delta \Omega}{2})^2 + F^2}, \\ \nonumber
E_{\rm s}  & = & \Omega_0 + \Omega_1- \frac{\Delta \Omega}{2} +
\sqrt{(\frac{\Delta \Omega}{2})^2 + F^2},
\end{eqnarray}
$\Omega_0$ denotes the ground state energy of the
system in which both the quantum dots are not occupied by excitons
while $\Omega_1$ denotes a higher  energy level of the excitonic
qubit system. The energy difference between the eigenstates is 
$\sqrt{{\Delta \Omega}^2 + (2 F)^2}$ and can be studied as
a function of time. In the absence of any decoherence
process, the excitonic qubit oscillates coherently between the
two dots with the Rabi frequency $E_{\rm s} - E_{\rm as}$.
 The polar  angle $\beta$ in  the Bloch sphere
representation of a qubit is related to $\Delta \Omega$ and $F$ via
\begin{equation}
\label{eq:angle}
\tan \beta = \frac{2 F}{\Delta \Omega}
\end{equation}

\section{\label{Forster} Interdot F\"orster amplitude term $F$}

The magnitude of the interdot F\"orster amplitude $F$  is 
given by the  matrix element 
\begin{eqnarray}
\label{eq:Forstermat}
F &=& \bra{\bf R}  \hat H_{F} \ket{\bf L} \\ 
\nonumber &=&  \sum_{\bf r_{\rm a}, r_{\rm b}}\; 
 \bra{g, {\bf R_{\rm a} }; f, {\bf R_{\rm b} }}
U(| {\bf r_{\rm a} - r_{\rm b}}|) 
\ket{f, {\bf R_{\rm a} }; g, {\bf R_{\rm b} }}  \\
\nonumber &+& \sum_{\bf r_{\rm a}, r_{\rm b}}\; 
 \bra{g, {\bf R_{\rm a} }; f, {\bf R_{\rm b} }}
U(| {\bf r_{\rm a} - r_{\rm b}}|) 
\ket{g, {\bf R_{\rm b} }; f, {\bf R_{\rm a} }}
\; 
\end{eqnarray}
where 
\[
U(|{\bf r_{{\rm a}}}-{\bf r_{{\rm b}}}|)=
 \frac{e^2}{\epsilon |{\bf r_{{\rm
a}}}-{\bf r_{{\rm b}}}|},
\]
and ${\bf r_{\rm a}}$ and  ${\bf r_{\rm a}}$ are position vectors 
of the charge carriers with origins at the center of the quantum dots
at ${\bf R_{\rm a}}$ and ${\bf R_{\rm b} }$ respectively.
The background dielectric constant, $\epsilon = 
\epsilon_0 \; \epsilon_r$ where the relative permittivity,
 $\epsilon_r$ is assumed to be independent of the location
of charge carriers. In the two-particle interaction matrix element
\[
\bra{g, {\bf R_{\rm a} }; f, {\bf R_{\rm b} }}
U(| {\bf r_{\rm a} - r_{\rm b}}|) 
\ket{f, {\bf R_{\rm a} }; g, {\bf R_{\rm b} }},
\]
the states  to the right of the scattering potential 
$U(| {\bf r_{{\rm a}}}-{\bf r_{{\rm b}}}| )$ represent 
the initial states while those to the left represent the  final
 scattered states. The first matrix term in Eq. \ref{eq:Forstermat} 
is due to Coulomb interaction in which the
 excited electron in its initial state $f$
in the  quantum dot at ${\bf R_{\rm a} }$ and a hole in the 
$g$th  state in the  quantum dot at ${\bf R_{\rm b} }$ 
get scattered through the potential $U$ to final states in 
which the electron and hole remain in the same excited 
states but their positions are exchanged. The
net effect can be viewed as a tunnelling process in which 
the exciton gets physically transferred from one quantum dot to
 another. The second matrix element in Eq. \ref{eq:Forstermat}  
represents the scattering of an electron in its initial state 
$f$ in the  quantum dot at ${\bf R_{\rm a} }$ and a hole in the 
$g$th  state in the  quantum dot at ${\bf R_{\rm b} }$ through 
the potential $U$ such that they remain on their respective
 quantum dot sites but their excited states are exchanged.  
This second process can be viewed 
as one in which the excited electron in one quantum dot
recombines with the hole in the ground state and the
liberated energy is transferred to excite 
the  electron-hole pair in a neighbouring quantum
dot. We assume that the quantum dots are separated by a distance,
 ${\rm W} = |{\bf R_{\rm a}} - {\bf R_{\rm b}}|$ such that 
${\rm W} \gg |{\bf r_{{\rm a}}}-{\bf r_{{\rm b}}}|$ so that  tunneling
effects are negligible.
The  interdot F\"orster energy transfer process  is then determined
by the second  matrix element in Eq. \ref{eq:Forstermat}.

Substituting expressions for $\ket{\bf R_{\rm a}}$ and
$\ket{\bf R_{\rm b}}$ (see Eq. \ref{eq:istate}) into
Eq. \ref{eq:Forstermat} and  employing the F\"orster 
multipole expansion of the Coulomb interaction we obtain:
\begin{eqnarray}
\label{eq:Forsterappr}
\bra{\bf R}  \hat H_{F} \ket{\bf L}
 &=&   \frac {\nu_0^2}{4 \pi \epsilon | {\bf R_{\rm a} - R_{\rm b}}|^3}
 \sum_{\bf r_e,r_h}\sum_{\bf r_e^\prime,r_h^\prime}  \; 
\Psi({\bf R_{\rm a},r_{e\parallel},r_{h\parallel}}) \;
\Psi^*({\bf R_{\rm b},r_{e\parallel}^\prime ,r_{h\parallel}^\prime})
\\ \nonumber  &\times& \varphi_e(z_e) \;
 \varphi_h(z_h)\; \varphi^* _e(z^\prime_e) 
\; \varphi^* _h(z^\prime_h) \; \delta_{\bf r_e,r_h}\; 
\delta_{\bf r_e^\prime,r_h^\prime}
\\ \nonumber  &\times &
\left [{\bf \mu_a \cdot \mu_b} - 
3 ({\bf  \mu_a \cdot R_d}) ({\bf  \mu_b \cdot R_d}) \right ] 
\end{eqnarray}
where 
\begin{equation}
{\bf R_{\rm d}} = \frac{\left ( {\bf R_{\rm a}} - {\bf R_{\rm b}} \right )}
{|{\bf R_{\rm a}} - {\bf R_{\rm b}}|},
\nonumber
\label{eq:vect}
\end{equation}
and 
\begin{equation}
\nonumber
\label{dipole}
{\bf  \mu_{a(b)}} = \int u_c({\bf{r}})\; ({\bf r_{\rm a(b)} - r}) 
\; u_v({\bf{r}}) \; d {\bf r}
\end{equation}
where $u_c({\bf{r}})$ and $u_v({\bf{r}})$ are the respective 
periodic components of the bulk Bloch functions of the 
electron and hole. We assume that the Bloch functions are 
identical in both the quantum dot and barrier materials.

Using  Eqs. \ref{eq:exciwavefnlat} -  \ref{eq:exciwavefnvert} in
Eq. \ref{eq:Forsterappr}, we  obtain:
\begin{eqnarray}
\label{eq:Forster}
F({\rm W})
 & = &  \bra{\bf R}  \hat H_{F} \ket{\bf L} \\ \nonumber
& = & \sqrt{\frac{11}{8}} \frac{\mu^2}{\epsilon {\rm W}^3}  \; 
\left ( \frac {2 {\rm l_e} {\rm l_h}}{ {\rm l_e}^2 +{\rm l_h}^2} \right )
\left ( \frac {2 {\rm l_{ze}} {\rm l_{zh}}}
{ {\rm l_{ze}}^2 +{\rm l_{zh}}^2} \right )^{1/2}
\end{eqnarray}
where $\mu = \mu_a \simeq \mu_b$ and ${\rm W} = 
|{\bf R_{\rm a}} - {\bf R_{\rm b}}|$. From Eq. 
\ref{eq:statesEigen} one obtains
\begin{eqnarray}
\label{eq:Fterms}
 \bra{\chi_{\rm s}} \hat H_F  \ket{\chi_{\rm s} }
& = &  - \bra{\chi_{\rm as}} \hat H_F  \ket{\chi_{\rm as} } =
\sin \beta \; F({\rm W})  \\ \nonumber
\bra{\chi_{\rm as}} \hat H_F  \ket{\chi_{\rm s} }
& = & - \cos \beta \; F({\rm W})
\end{eqnarray}
In view of the fact that $F$ is key to the entanglement and the
formation of entangled states, $\ket{\chi_{\rm s}}$ 
and  $\ket{\chi_{\rm as}}$, we expect the inevitable 
 loss of energy to lattice vibrations coupled with the excitation transfer 
to lead to decoherence and subsequent degradation of qubit states. We will
examine this important process in greater detail in the next section.

\section{\label{phonon} Excitonic qubit-Phonon Interaction}
%\label{sec:phonon}
We consider  the  exciton state at each  quantum 
dot to be  coupled to a continuum of acoustic phonons  via both 
deformation potential and piezolelectric coupling.
 The acoustic phonons are modelled as bulk modes which 
is a valid approximation for quantum dots fabricated
in barrier materials with almost similar lattice properties
such as (In,Ga)As quantum dots with a (Ga,Al)As  
barrier.  The total Hamiltonian of
a system of   F\"orster  coupled quantum dot 
interacting with phonons is given by 
\begin{eqnarray}
\label{eq:envqubit}
\hat H_{\rm qb}^{\rm env} = \hat H_{\rm ex-qb} + \hat H^{\rm ph} 
 + \hat H_{\rm ex-qb}^{\rm DP} \\ \nonumber + 
\hat H_{{\rm ex-qb}, \lambda}^{\rm Piez} + 
\hat H_{\rm F} + \hat H_{\rm F}^{\rm ph}
\end{eqnarray}
where $\hat H_{\rm ex-qb} $ is given by Eq. \ref{eq:Hamqubit} and 
the matrix elements of $\hat H_{\rm F}$ is defined by the 
F\"orster amplitude term $F$ in Eq. \ref{eq:Forster}.
$\hat H^{\rm ph}$ is the phonon reservoir
\begin{equation}
\nonumber
\label{eq:phonon}
\hat H_{\rm ph} = \sum_{\bf q} \hbar \omega_{_{{\bf q} {\lambda}}}\,
b_{\lambda}^{\dagger}({\bf q})\,b_{\lambda}(\bf{q}),
\end{equation}
where  $b_{\lambda}^{\dagger}(\bf{q})\,$ and $b_{\lambda}(\bf{q})\,$
are the respective  creation and  annihilation 
operator of a $\lambda-$mode phonon with 
wave vector ${\bf q}$. The $\lambda$-mode is denoted ${\rm LA}$ 
for longitudinal acoustic phonons and ${\rm TA}$ 
for transverse acoustic  phonons. The acoustic phonon energy spectrum
 is determined by  the dispersion relation 
$\omega_{_{{\bf q} {\rm LA}}} = \upsilon_{_{{\rm LA}}} |\bf{q}| $ 
for the longitudinal mode and $\omega_{_{{\bf q} {\rm TA}}} 
= \upsilon_{_{{\rm TA}}} |\bf{q}|$ for the transverse mode, with 
$\upsilon_{_{{\rm LA}}}$ and $\upsilon_{_{{\rm TA}}}$ 
denoting the corresponding sound velocities. 

$\hat H_{\rm ex-qb}^{\rm DP}$ is the Hamiltonian for
excitonic qubit-phonon  interaction  via 
 deformation potential coupling and is 
linear in terms of phonon creation and annihilation operators
\begin{eqnarray}
\label{eq:exphDP}
\hat H_{\rm ex-qb}^{\rm DP} = 
\sum_{\lambda,{\bf q}}  \sqrt{ \frac{\hbar \; |{\bf q}|^2} 
{2\rho\, V\, \omega_{{\bf q} \, \lambda}}}\;
\left [ {\rm M_r} \,  \sigma_{x} + {\rm M_p} \, \sigma_{z} \right ]
\\
\nonumber
\times \left ( b^\dagger_{\lambda}({\bf -q}) +
b_{\lambda}({\bf q})\right ) \; 
\ket{n_{{\bf q} \, \lambda}} \bra{n_{{\bf q} \, \lambda}}  
\end{eqnarray}
where $\ket{n_{{\bf q} \, \lambda}}$ 
denotes the occupation number of $\lambda-$mode phonon with
wave vector,${\bf q}$. $V$ is the crystal volume
and $\rho$ is the  mass density of the material system.
$D_c$ and $D_v$ are the respective deformation potential
 constants for the conduction and valence bands.
The terms ${\rm M_r}$  and ${\rm M_p}$ are written as
\begin{eqnarray}
\label{eq:Mtermr}
{\rm M_r}   = \bra{\chi_{\rm as}; n_{{\bf q} \, \lambda}\; \pm \; 1} 
\left (D_c e^{\rm i\, {\bf q.r_e}}- D_v  e^{\rm i\, {\bf q.r_h}}\right )
 \ket{\chi_{\rm s}; n_{{\bf q} \, \lambda} },
\end{eqnarray}
and
\begin{eqnarray}
\label{eq:Mtermp}
{\rm M_p}  & = & \bra{\chi_{\rm s}; n_{{\bf q} \, \lambda} \; \pm \; 1}
 \left ( D_c e^{\rm i\, {\bf q.r_e}}- D_v  e^{\rm i\, {\bf q.r_h}} \right )
 \ket{\chi_{\rm s}; n_{{\bf q} \, \lambda} } \\ \nonumber
& - &  \bra{\chi_{\rm as}; n_{{\bf q} \, \lambda} \; \pm \; 1} 
\left ( D_c e^{\rm i\, {\bf q.r_e}}- D_v  e^{\rm i\, {\bf q.r_h}} \right )
 \ket{\chi_{\rm as}; n_{{\bf q} \, \lambda} } 
\end{eqnarray}
The ${\rm M_r}$  term describes a decoherence process in which the
qubit state  $\ket{\chi_{\rm s}}$ relaxes to the qubit state  
$\ket{\chi_{\rm as}}$ due to mediation by phonons.
The  ${\rm M_p}$ term describes a second type of 
decoherence process in which a  shift occurs in the energy difference 
between the  two qubit states resulting in pure dephasing of the entangled system.  

A similar expression as in Eq. \ref{eq:exphDP} can be obtained 
for the Hamiltonian $\hat H_{{\rm ex-qb}, \lambda}^{{\rm Piez}}$ describing 
exciton-phonon interaction via piezoelectric coupling
\begin{eqnarray}
\label{eq:exphPiez}
\hat H_{{\rm ex-qb}, \lambda}^{{\rm Piez}}  &=&  
\sum_{\lambda,{\bf q}}  \;
 \frac{8 \pi e e_{14}}{\epsilon_0 \epsilon_r |{\bf q}|^2}
\sqrt{ \frac{\hbar} {2\rho\, V\, \omega_{{\bf q} \, \lambda}}}\;
\left(\xi_{x,\lambda} q_y q_z + 
\xi_{y,\lambda} q_x q_z +  \xi_{z,\lambda} q_x q_y \right )
\\ \nonumber & \times &
\left [ {\rm N_r} \,  \sigma_{x} + {\rm N_p} \, \sigma_{z} \right ]
\left ( b^\dagger_{\lambda}({\bf -q}) +
b_{\lambda}({\bf q})\right ) \; 
\ket{n_{{\bf q} \, \lambda}} \bra{n_{{\bf q} \, \lambda}}  
\end{eqnarray}
 where the relative permittivity,
 $\epsilon_r$ is assumed to be unaffected by 
the  contribution from strain fields associated with
acoustic phonon modes. $e_{14}$ is the piezoelectric constant and
$\xi_{_{i,\lambda}}$ is the  unit vector of polarization of the 
$\lambda$-phonon  along the $i$-direction. 
Excitonic interactions with phonons due to piezoelectric coupling 
are  highly anisotropic in nature \cite{thilaM} and the form
of $\hat H_{{\rm ex-qb}, \lambda}^{{\rm Piez}}$ depends
on the choice of polarization components and the modes
associated with $\lambda$. Further details of the dependency
on ${\rm LA}$, ${\rm TA1}$ and ${\rm TA2}$ modes are given in the
next section. ${\rm N_r}$  and ${\rm N_p}$ are analogous 
to ${\rm M_r}$ and ${\rm M_p}$ given in  
 Eqs. \ref{eq:Mtermr} and \ref{eq:Mtermp}
respectively
\begin{eqnarray}
\label{eq:Ntermr}
{\rm N_r}   = \bra{\chi_{\rm as}; n_{{\bf q}\; \lambda}\pm 1} 
\left (e^{\rm i\, {\bf q.r_e}}- e^{\rm i\, {\bf q.r_h}} \right )
 \ket{\chi_{\rm s}; n_{{\bf q}\;\lambda} },
\end{eqnarray}
and 
\begin{eqnarray}
\label{eq:Ntermp}
{\rm N_p}  & = & \bra{\chi_{\rm s}; n_{{\bf q} \, \lambda}\pm 1} 
\left (e^{\rm i\, {\bf q.r_e}}-  e^{\rm i\, {\bf q.r_h}} \right )
 \ket{\chi_{\rm s}; n_{{\bf q} \, \lambda} } \\ \nonumber
& - &  \bra{\chi_{\rm as}; n_{{\bf q} \, \lambda} \pm 1} 
\left (e^{\rm i\, {\bf q.r_e}}-  e^{\rm i\, {\bf q.r_h}} \right )
 \ket{\chi_{\rm as}; n_{{\bf q} \, \lambda} } 
\end{eqnarray}
Unlike the ${\rm M_r}$ and ${\rm M_r}$ terms,  
both  ${\rm N_r}$ and ${\rm N_p}$ vanish 
as $ {\bf q} \rightarrow 0$  due to exact 
 cancellation of electron and hole form factors as 
the  piezoelectric coupling constant remains the same for
electrons and holes.

We consider the Hamiltonian $\hat H_{\rm F}^{\rm ph}$
in Eq. \ref{eq:envqubit} to be associated with 
a interdot F\"orster energy transfer process that  proceeds
 via a one-phonon assisted transfer of excitonic qubits. 
A one-phonon-assisted decoherence of qubit states is
 more likely at lower temperatures while  multiphonon
phonon processes are expected to dominate at higher temperatures.
We express $\hat H_{\rm F}^{\rm ph}$ as 
arising from second-order process involving two Hamiltonians
\begin{equation}
\label{eq:2ndorder}
\hat H_{\rm F}^{\rm ph}
=  \; \hat H_{\rm ex-qb}^{\rm X} \oplus 
H_{\rm F}
\end{equation}
where  ${\rm X = DP (Piez)}$ for coupling via deformation potential 
(piezoelectric fields). The relaxation of qubit states occurs 
via several routes
\begin{eqnarray}
\label{eq:pathrelax1}
 \ket{\chi_{\rm s}; n_{{\bf q} \;\lambda}} &&
 \stackrel{\hat H_{\rm ex-qb}^{\rm X}}{\longrightarrow} 
\ket{\chi_{\rm as}; n_{{\bf q} \;\lambda} \pm 1} 
\stackrel{\hat H_{\rm F}}{\longrightarrow}
\ket{\chi_{\rm as}; n_{{\bf q}\;\lambda} \pm 1},  \\ \nonumber
\label{eq:pathrelax2}
 \ket{\chi_{\rm s}; n_{{\bf q}\;\lambda}} &&
 \stackrel{\hat H_{\rm ex-qb}^{\rm X}}{\longrightarrow} 
\ket{\chi_{\rm s}; n_{{\bf q}\;\lambda} \pm 1} 
\stackrel{\hat H_{\rm F}}{\longrightarrow}
\ket{\chi_{\rm as}; n_{{\bf q}\;\lambda} \pm 1}
, \\ \label{eq:pathrelax3}
 \ket{\chi_{\rm s}; n_{{\bf q}\;\lambda}} &&
 \stackrel{\hat H_{\rm F}}{\longrightarrow} 
\ket{\chi_{\rm s}; n_{{\bf q}\;\lambda}} 
\stackrel{\hat H_{\rm ex-qb}^{\rm X}}{\longrightarrow}
\ket{\chi_{\rm as}; n_{{\bf q}\;\lambda} \pm 1},\\ \nonumber
\label{eq:pathrelax4} 
 \ket{\chi_{\rm s}; n_{{\bf q}\;\lambda}} &&
 \stackrel{\hat H_{\rm F}}{\longrightarrow} 
\ket{\chi_{\rm as}; n_{{\bf q}\;\lambda} } 
\stackrel{\hat H_{\rm ex-qb}^{\rm X}}{\longrightarrow}
\ket{\chi_{\rm as}; n_{{\bf q}\;\lambda} \pm 1}
\end{eqnarray}
Likewise,  pure dephasing of   excitonic qubit state
 $\ket{\chi_{\rm s}; n_{\bf q}}$  can occur via
\begin{eqnarray}
\nonumber
\label{eq:pathphase1}
 \ket{\chi_{\rm s}; n_{{\bf q}\;\lambda}} &&
 \stackrel{\hat H_{\rm ex-qb}^{\rm X}}{\longrightarrow} 
\ket{\chi_{\rm as}; n_{{\bf q}\;\lambda} \pm 1} 
\stackrel{\hat H_{\rm F}}{\longrightarrow}
\ket{\chi_{\rm s}; n_{{\bf q}\;\lambda} \pm 1},  \\ 
\nonumber
\label{eq:pathphase2}
 \ket{\chi_{\rm s}; n_{{\bf q}\;\lambda}} &&
 \stackrel{\hat H_{\rm ex-qb}^{\rm X}}{\longrightarrow}
\ket{\chi_{\rm s}; n_{{\bf q}\;\lambda} \pm 1} 
\stackrel{\hat H_{\rm F}}{\longrightarrow}
\ket{\chi_{\rm s}; n_{{\bf q}\;\lambda} \pm 1}
, \\  \nonumber
\label{eq:pathphase3}
 \ket{\chi_{\rm s}; n_{{\bf q}\;\lambda}}  &&
 \stackrel{\hat H_{\rm F}}{\longrightarrow}
\ket{\chi_{\rm s}; n_{{\bf q}\;\lambda} } 
\stackrel{\hat H_{\rm ex-qb}^{\rm X}}{\longrightarrow}
\ket{\chi_{\rm s}; n_{{\bf q}\;\lambda} \pm 1}
,\\
\label{eq:pathphase4} 
\ket{\chi_{\rm s}; n_{{\bf q}\;\lambda}} &&
 \stackrel{\hat H_{\rm F}}{\longrightarrow} 
\ket{\chi_{\rm as}; n_{{\bf q}\;\lambda} } 
\stackrel{\hat H_{\rm ex-qb}^{\rm X}}{\longrightarrow}
\ket{\chi_{\rm s}; n_{{\bf q}\;\lambda} \pm 1}
\end{eqnarray}
Pure dephasing of  the  qubit state
 $\ket{\chi_{\rm as}; n_{\bf q}}$  occurs
similarly as shown above for the 
$\ket{\chi_{\rm s}; n_{\bf q}}$ state.

\section{ \label{sec:matrix} Evaluation of Matrix Elements}
Using Eqs. \ref{eq:istate}, \ref{eq:exciwavefnlat}, \ref{eq:fouriertransform}
and \ref{eq:exciwavefnvert} in Eq. \ref{eq:exphDP},  we obtain 
 explicit forms for the following matrix elements
\begin{eqnarray}
\label{eq:PmatrixDPa} 
\bra{{\bf L}; n_{{\bf q}\;\lambda} \pm 1}
 \hat H_{\rm ex-qb}^{\rm DP}  \ket{{\bf L; n_{{\bf q} \,\lambda} }}
& =  & \Sigma_D(q_\parallel,q_z)\; e^{- {\rm i} {\bf q \cdot R_{\rm a} }}
\\ \label{eq:PmatrixDPb}
 \bra{{\bf L}; n_{{\bf q}\;\lambda} \pm 1}
 \hat H_{\rm ex-qb}^{\rm DP}  \ket{{\bf R; n_{{\bf q} \,\lambda} }}
& = & \Sigma_D(q_\parallel,q_z)\;
e^{- {\rm i} {\bf q \cdot \frac{(R_{\rm a} + R_{\rm a})}{2}}} 
 e^{-\frac{{\rm W}^2}{4{\rm L_R}^2} }
\end{eqnarray}
where ${\rm L_R}$ is defined below Eq. \ref{eq:exciwavefnlat}
 and a similar expression as in Eq. \ref{eq:PmatrixDPa} 
can be obtained for $\bra{{\bf R}; n_{{\bf q}\;\lambda} \pm 1}
 \hat H_{\rm ex-qb}^{\rm DP}  \ket{{\bf R; n_{{\bf q} \,\lambda} }}$. 
The function $\Sigma_D(q_\parallel,q_z)$ which is given by
\begin{eqnarray}
\nonumber
\label{eq:funcDP}
\Xi_D(q_\parallel,q_z) =  \sqrt{\frac{\hbar \; |{\bf q}|^2}
 {2\rho\, V\, \omega_{{\bf q}\; \lambda}}}
 e^{-\frac{1}{4}{\rm L_R}^2 {q_\parallel}^2}
 \left [D_c \; e^{-\frac{1}{4}{\rm l_{ze}}^2 q_z^2}\; 
e^{-\frac{1}{4}{\rm l_e}^2 q_\parallel^2} - 
D_v  e^{-\frac{1}{4}{\rm l_{zh}}^2 q_z^2}\; 
e^{-\frac{1}{4}{\rm l_h}^2 q_\parallel^2}\right ] \; 
\end{eqnarray}
differs from the commonly used form \cite{Fed} as we consider
the non-isotropic propagation of phonon here.
From Eqs. \ref{eq:statesEigen}, \ref{eq:PmatrixDPa} 
and \ref{eq:PmatrixDPb} we obtain
\begin{eqnarray}
\label{eq:PmatrixDPr1}
\bra{\chi_{\rm as}; n_{{\bf q}\;\lambda} \pm 1}
 \hat H_{\rm ex-qb}^{\rm DP}  \ket{\chi_{\rm s}; n_{{\bf q} \,\lambda} }
&=&  \frac{1}{2}\sin(\beta /2)
 \left ( e^{- {\rm i} {\bf q \cdot R_{\rm a}}} - 
e^{- {\rm i} {\bf q \cdot R_{\rm b}}} \right) \Sigma_D(q_\parallel,q_z)
\\ \nonumber &-& \cos(\beta /2)
e^{-\frac{{\rm W}^2}{4{\rm L_R}^2} }\Sigma_D(q_\parallel,q_z)
\end{eqnarray}
The first term in  Eq. \ref{eq:PmatrixDPr1} contains the coherence factor 
$\left ( e^{- {\rm i} {\bf q \cdot R_{\rm a}}} - 
e^{- {\rm i} {\bf q \cdot R_{\rm b}}} \right)$ 
which remains effective over large distances between the 
quantum dots. The second term involves the   factor
 $e^{-\frac{{\rm D}^2}{4{\rm L_R}^2} }$ which is
short-ranged as it depends on the overlap integral of
 the exciton wavefunctions in quantum dots located at 
different sites. By integrating the coherence factor over the 
polar angle $\phi$, we obtain 
\begin{eqnarray}
\label{eq:PmatrixDPr2}
|\bra{\chi_{\rm as}; n_{{\bf q}\;\lambda} \pm 1}
 \hat H_{\rm ex-qb}^{\rm DP}  \ket{\chi_{\rm s}; n_{{\bf q} \,\lambda} }|^2
&=&  2 \pi \Sigma_D^2(q_\parallel,q_z)
 \left [\frac{1}{2} \sin^2(\beta) (1-J_0(q_\parallel {\rm W}))
+ \cos^2(\beta)e^{-\frac{{\rm W}^2}{2{\rm L_R}^2} }\right ]
\end{eqnarray}
where $J_0(q_\parallel {\rm W})$ is the zeroth-order Bessel function. 
In order to obtain a tractable form of Eq. \ref{eq:PmatrixDPr2} we 
approximate $J_0(q \sin(\theta) {\rm W})$ by taking the angular average 
\begin{eqnarray}
\nonumber
\label{eq:approx}
J_0(q \sin(\theta) {\rm W}) && \approx 
\frac{\int \int J_0(q \sin(\theta) {\rm W}) \sin(\theta) d\theta d \phi}
{\int \int \sin(\theta) d\theta d \phi} \\ \nonumber
&& = \frac{\sin(q {\rm W})}{q {\rm W}}
\end{eqnarray}
For large separations between the quantum dots, we approximate 
\begin{eqnarray}
\nonumber
\label{eq:PmatrixDPrL}
|\bra{\chi_{\rm as}; n_{{\bf q}\;\lambda} \pm 1}
 \hat H_{\rm ex-qb}^{\rm DP}  \ket{\chi_{\rm s}; n_{{\bf q} \,\lambda} }|^2
& \approx &   \pi \Sigma_D^2(q_\parallel,q_z)
\sin^2(\beta) (1-\frac{\sin(q {\rm W})}{q {\rm W}})
\end{eqnarray}
Similar expressions as in Eq. \ref{eq:PmatrixDPr2} but 
with slight variations in the dependency on polar  
angle $\beta$  can  be obtained for 
$|\bra{\chi_{\rm as}; n_{{\bf q}\;\lambda} \pm 1}
 \hat H_{\rm ex-qb}^{\rm DP}  \ket{\chi_{\rm as}; n_{{\bf q} \,\lambda} }|^2$ and
$|\bra{\chi_{\rm s}; n_{{\bf q}\;\lambda} \pm 1}
 \hat H_{\rm ex-qb}^{\rm DP}  \ket{\chi_{\rm s}; n_{{\bf q} \,\lambda} }|^2$.
 These expressions will be used in  Eqs. \ref{eq:Mtermr} 
and \ref{eq:Mtermp} to evaluate the rates at which
the excitonic qubits lose their decoherence via relaxation 
and dephasing in the following section.

In the case of qubit-phonon interaction via piezoelectric coupling,
we use a choice of phonon polarization components with respect to the cubic
crystallographic axes of the zinc-blende type crystal. Substituting
Eqs. \ref{eq:istate}, \ref{eq:exciwavefnlat}, \ref{eq:fouriertransform}
and \ref{eq:exciwavefnvert} into Eq. \ref{eq:exphPiez},
we get  explicit forms for the following matrix elements 
\begin{eqnarray}
\label{eq:PmatrixPia} 
\bra{{\bf L}; n_{{\bf q}\;\lambda} \pm 1}
 \hat H_{{\rm ex-qb},\lambda}^{\rm Piez}  \ket{{\bf L; n_{{\bf q} \,\lambda} }}
& = & \Sigma_P(q_\parallel,q_z,\lambda)\; e^{-{\bf q_\parallel \cdot R_{\rm a} }}
\\ \label{eq:PmatrixPib}
 \bra{{\bf L}; n_{{\bf q}\;\lambda} \pm 1}
 \hat H_{{\rm ex-qb},\lambda}^{\rm Piez}  \ket{{\bf R; n_{{\bf q} \,\lambda} }}
& = & \Sigma_P(q_\parallel,q_z,\lambda)\;
e^{-{\bf q_\parallel \cdot \frac{(R_{\rm a} + R_{\rm a})}{2}}} 
 e^{-\frac{{\rm W}^2}{4{\rm L_R}^2} }
\end{eqnarray}
where the function $\Sigma_P(q_\parallel,q_z,\lambda)$ is defined as
\begin{eqnarray}
\label{eq:funcPi}
\Sigma_P(q_\parallel,q_z,\lambda) = \frac{4 \pi e e_{14}}{\epsilon_0 \epsilon_r}
\sqrt{ \frac{\hbar} {2\rho\, V\, \upsilon_{\lambda} q}}\;
{\rm A}_{\lambda}(\theta,\phi) \; e^{-\frac{1}{4}{\rm L_R}^2 {q_\parallel}^2}
\left [e^{-\frac{1}{4}{\rm l_{ze}}^2 q_z^2}\;
 e^{-\frac{1}{4}{\rm l_e}^2 q_\parallel^2} - 
e^{-\frac{1}{4}{\rm l_{zh}}^2 q_z^2}\; 
e^{-\frac{1}{4}{\rm l_h}^2 q_\parallel^2}\right ]
\end{eqnarray}
The anisotropy factor ${\rm A}_{\lambda}(\theta,\phi)$ is given by
\begin{eqnarray}
\label{eq:anfactorLA}
{\rm A}_{{\rm LA}}(\theta,\phi)  &=& \frac{3}{4} \sin 2\theta \sin \theta
\sin 2\phi \;,
\\ \label{eq:anfactorTA1}
{\rm A}_{{\rm TA1}}(\theta,\phi)  &=& \frac{1}{8} ( \sin \theta - 
 \sin 3\theta)\sin 2\phi \;,
\\ \label{eq:anfactorTA2}
{\rm A}_{{\rm TA2}}(\theta,\phi)  &=& \frac{1}{2} \sin 2\theta 
\cos 2\phi \;
\end{eqnarray}
The ${\rm TA1}$ and ${\rm TA2}$ modes   correspond to the two possible
polarized direction of the transverse phonon. 
Calculations of the square amplitude term $|\bra{\chi_{\rm as}; 
n_{{\bf q}\;\lambda} \pm 1}  \hat H_{\rm ex-qb,\lambda}^{\rm Piez} 
\ket{\chi_{\rm s}; n_{{\bf q} \,\lambda} }|^2$ where
$\lambda = {\rm LA, TA1, TA2}$ are slightly involved  due
 to the presence of the polar angle $\phi$ in Eq. \ref{eq:anfactorLA} - 
\ref{eq:anfactorTA2}. For $\lambda = {\rm LA}$,   we obtain 
\begin{eqnarray}
\label{eq:PmatrixPi2}
|\bra{\chi_{\rm as}; n_{{\bf q}\;{\rm LA} \pm 1}}
 \hat H_{\rm ex-qb,LA}^{\rm Piez}  \ket{\chi_{\rm s}; n_{{\bf q} \,{\rm LA} }}|^2
= \frac{\pi}{4}\Sigma_P^2(q_\parallel,q_z,{\rm LA}) 
\left [\sin^2(\beta) (1-\frac{\sin(q {\rm W})}{q {\rm W}})
+ 9 \cos^2(\beta)e^{-\frac{{\rm W}^2}{2{\rm L_R}^2} }\right ]
\end{eqnarray}
where $\Sigma_P(q_\parallel,q_z,{\rm LA})$ can be obtain using
Eqs. \ref{eq:funcPi} and \ref{eq:anfactorLA}.
Similar expressions as in Eq. \ref{eq:PmatrixPi2} 
 can be obtained for  the  ${\rm TA1}$ and ${\rm TA2}$ 
modes  of the transverse phonon using Eqs. \ref{eq:anfactorTA1} 
and \ref{eq:anfactorTA2}. 

\section{\label{Decor}  Decoherence rates of excitonic qubits}

The decoherence rates associated with relaxation and pure dephasing
processes are  calculated using  
\begin{eqnarray}
\label{eq:rate}
\frac{1}{\tau_{_{{\rm X},\lambda}}} 
 =  \frac{2 \pi}{ \hbar}\, \sum _{q_\parallel,q_z} \, 
{| \langle f| \hat H_{{\rm int}}|i \rangle|}^2
\, (N_{{\bf q}, \lambda} +1) \; \delta(\Delta 
E \pm \hbar \omega_{{\bf q}\,\lambda})
\end{eqnarray}
where  $|i \rangle$  and $|f \rangle$ denote the initial
and final states which are functions of $\ket{\chi_{\rm as}}$
and  $\ket{\chi_{\rm s}}$  (see  Eqs. \ref{eq:Mtermr},
\ref{eq:Mtermp}, \ref{eq:Ntermr} and \ref{eq:Ntermp}).
$\Delta E$ is the energy difference between the initial and 
final states which determines the wavevector of the emitted phonon.
 $H_{{\rm int}}$  is the interaction operator
which is substituted by $\hat H_{\rm ex-qb}^{\rm DP}$ 
or  $\hat H_{{\rm ex-qb},\lambda}^{\rm Piez}$ 
 depending on the type of phonon coupling. In the 
case of  decoherence associated with  F\"orster transfer process,
$H_{{\rm int}}$ is given by  $\hat H_{\rm F}^{\rm ph}$ as will be 
detailed in the next section.  $ N_{{\bf q} \, \lambda}$  is the thermalised 
average number of phonons at the temperature $T$, and is 
given by the Bose-Einstein distribution 
$ N_{{\bf q} \, \lambda} = \exp (\hbar \omega_{{\bf q}\,\lambda}/k_B T)$
where $k_B$ is the Boltzmann constant.

From  Eqs. \ref{eq:Mtermr},  \ref{eq:Mtermp}, 
\ref{eq:PmatrixDPr2} and  \ref{eq:rate}
we obtain explicit expressions for the rate 
of relaxation ($1/\tau_{_{\rm DP}}^r$) and rate of dephasing
($1/\tau_{_{\rm DP}}^p $) of excitonic qubits
\begin{eqnarray}
\label{eq:raterDP}
1/\tau_{_{\rm DP}}^{r(p)} 
= \frac{(D_c-D_v)^2 q^3}{4 \sqrt{\pi} \hbar \rho \upsilon_{_{{\rm LA}}}^2}
\exp(-\frac{3}{4} q^2 {\rm l}^2) \frac{{\rm Erfi}(\Delta {\rm l})}
{\Delta {\rm l}} (N_{{\bf q}, \lambda} +1) S_{r(p)}(\beta)
 \end{eqnarray}
 where the functions $S_{r}(\beta)$ and  $S_{p}(\beta)$ are given by 
\begin{eqnarray}
\label{eq:Hexp}
S_{r}(\beta) && = 
\frac{1}{2} \sin^2(\beta) \left ( 1-\frac{\sin(q {\rm W})}{q {\rm W}} \right )
+ \cos^2(\beta)e^{-\frac{{\rm W}^2}{{\rm l}^2}} \\ \nonumber 
S_{p}(\beta) && = \frac{1}{2}\cos^2(\beta) 
\left(1-\frac{\sin(q {\rm W})}{q {\rm W}} \right )
+ \frac{1}{2}\sin^2(\beta)e^{-\frac{{\rm W}^2}{{\rm l}^2}}
\end{eqnarray}
To simplify the derivation of Eq. \ref{eq:raterDP}, we
 have used the assumptions that ${\rm l_e \approx l_h = l}$ and 
${\rm l_{ze} \approx l_{zh} = l_z}$. 
${\rm Erfi}(\Delta l)$ is the imaginary error function,  
$\Delta {\rm l} = q \sqrt{\frac{1}{2}(\frac{3}{2} {\rm l}^2-{\rm l_z}^2)}$ and 
$q = \frac{2 F({\rm W})}{\hbar \upsilon_{_{{\rm LA}}}} \sqrt{1+\gamma^2}$
where $\gamma = \frac{\Delta \Omega}{2 F({\rm W})}$
(see Eq. \ref{eq:Forster} for an analytical expression for $F({\rm W})$).
It can be noted that the  cut-off frequency in Eq. \ref{eq:raterDP}
 occurs at $\omega _{l} \sim {\upsilon_{_{{\rm LA}}}}/{ \rm l}$ 
which  is approximately the inverse phonon flight time through the
quantum dot.

%%%%%%%%%%%%%%%%%%
\begin{figure}[ht]
\includegraphics[width=11.4cm]{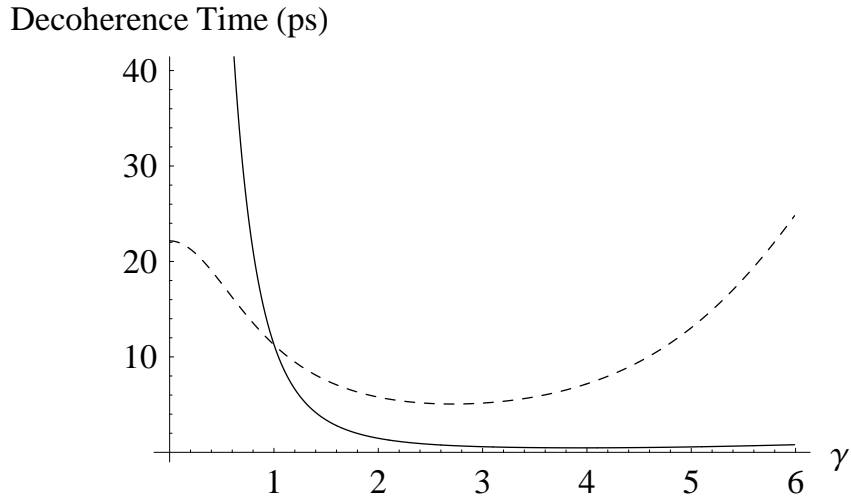}
\caption{Relaxation time $\tau_{_{\rm DP}}^r$ (dashed)
and dephasing time, $\tau_{_{\rm DP}}^p $ (full) as functions of 
$\gamma = \frac{\Delta \Omega}{2 F({\rm W})}$ at $T$ = 10 K, 
${\rm W}$ = 5 nm , ${\rm l_e \approx l_h = l}$ = 2.5 nm and 
${\rm l_{ze} \approx l_{zh} = l_z}$ =  2.5 nm.}
\label{fig:ratesDP}
\end{figure}
%%%%%%%%%%%%%%%%%%

%%%%%%%%%%%%%%%%%%
\begin{figure}[ht]
\includegraphics[width=11.4cm]{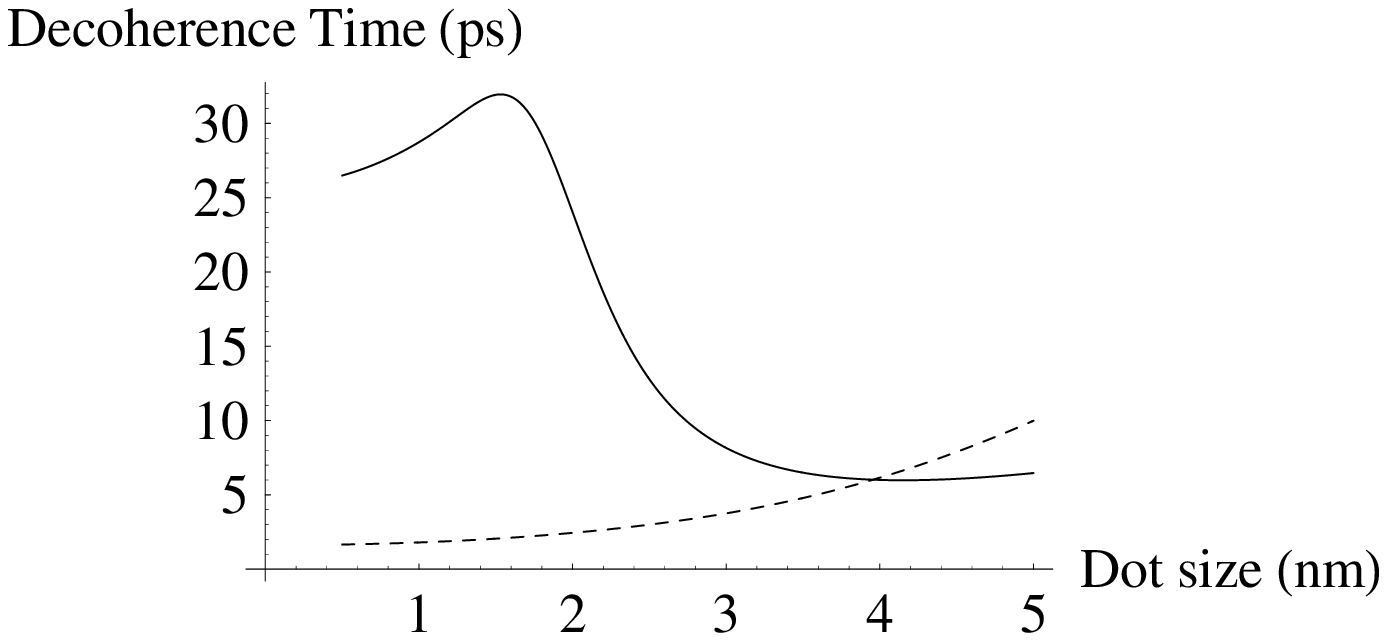}
\caption{Relaxation time, $\tau_{_{\rm DP}}^r$ (dashed)
and dephasing time $\tau_{_{\rm DP}}^p$ (full) as functions of 
dot size ${\rm l}$ at $\gamma$ = 0.25,  ${\rm W}$ = 4 nm, ${\rm l_z}$ = 1 nm and
$T$ = 10 K.}
\label{fig:DPsize}
\end{figure}
%%%%%%%%%%%%%%%%%%

%%%%%%%%%%%%%%%%%%
\begin{figure}[ht]
\includegraphics[width=11.4cm]{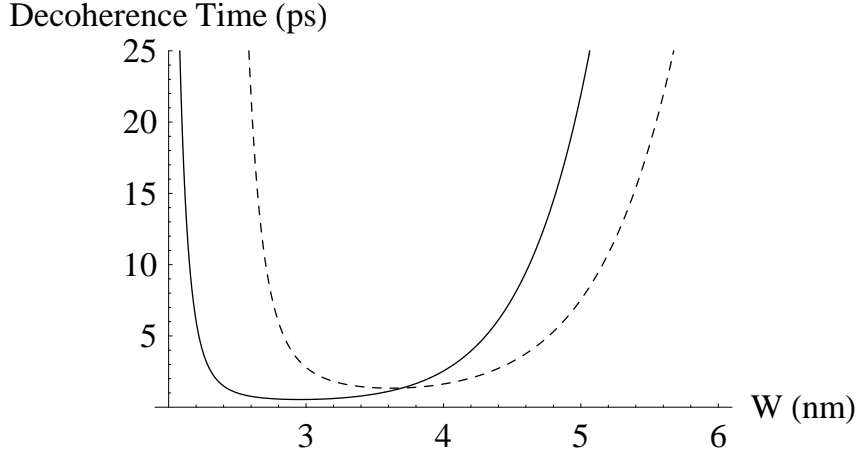}
\caption{Relaxation time $\tau_{_{\rm DP}}^r$
 as function of interdot distance $W$ at $\gamma$ = 1.5 (dashed) and
$\gamma$ = 0.1 (full) at  ${\rm l}$ = 2 nm, ${\rm l_z}$ = 1 nm and
$T$ = 10 K.}
\label{fig:DPW}
\end{figure}
%%%%%%%%%%%%%%%%%%

Using the parameters for the GaAs/AlGaAs material system \cite{thilaM}, 
we calculate the relaxation time $\tau_{_{\rm DP}}^r$ and
 dephasing time $\tau_{_{\rm DP}}^p $ as functions of 
$\gamma$  in  Fig.~\ref{fig:ratesDP}.
At  $\gamma = 0$, the qubit states are maximally entangled and
decoherence  is dominated by the relaxation process as one can expect 
from  two equally populated states. As $\gamma$ increases, the difference
in the population of the qubit states is enhanced and decoherence  becomes
increasingly  dominated by pure dephasing process. We note that the
 minimum in relaxation time $\tau_{_{\rm m}}^r$ is reached at $\gamma_m$
and  is given by $\tau_{_{\rm m}}^r \approx 
{\rm W}/2 \pi \upsilon_{_{\rm LA}}$ thus satisfying the condition where the
interdot separation matches the phonon wavevector. 
 At small values of $\gamma$, the quantum dot size parameter
 ${\rm l}$ has a stronger  influence on the dephasing process
as shown in Fig.~\ref{fig:DPsize} where the dephasing time decreases sharply after
${\rm l} \approx$ 2 nm while the relaxation time increases gradually mainly determined
by the exponential term in  Eq. \ref{eq:raterDP}.
 Fig.~\ref{fig:DPW} shows that at critical values of ${\rm W}$ (depending
on $\gamma$ and ${\rm l}$), the relaxation times can reach very small
 values in the order of several picoseconds.

In  the case of qubit-phonon interaction via piezoelectric coupling,
  we obtain the relaxation and dephasing rates  for $\lambda = {\rm LA}$,
using  Eqs. \ref{eq:Mtermr}, \ref{eq:Mtermp}, \ref{eq:PmatrixPi2} and  \ref{eq:rate}
\begin{eqnarray}
\label{eq:ratePiez}
1/\tau_{_{\rm  Piez}}^{r(p)}
= \frac{\pi^2 q e^2 e_{14}^2}{2 \epsilon^2
 \hbar \rho \upsilon_{_{{\rm LA}}}^2}\; \exp(-q^2 {\rm l}^2/2)
\left [1-\exp(-\frac{1}{4}(r^2-1) q^2 {\rm l}^2)\right]^2
 G_{_{{\rm LA}}}(\frac{q^2 {\rm l}^2 r^2}{2 \left(r^2+1\right)})
 (N_{{\bf q}, \lambda} +1)
 S_{r(p)}^\prime(\beta)
\end{eqnarray}
where $r = {\rm l_h/l_e}$ and we have used the assumption 
that ${\rm l = l_e  = l_{ze}}$ in order to obtain an analytical expression.
The function $G_{_{{\rm LA}}}(x)$ is given by
\begin{eqnarray}
\label{eq:gfun}
G_{_{{\rm LA}}}(x) && =
\frac{2 x + 15}{4 x^3}-\frac{ \sqrt{\pi } e^{-x} (
 4 x (x+3)+15) \text{Erfi}\left(\sqrt{x}\right)}{8 x^{7/2}}
\end{eqnarray}
and $S_{r}^\prime(\beta)$ and  $S_{p}^\prime(\beta)$ are given by 
\begin{eqnarray}
\label{eq:Hexpiez}
S_{r}^\prime(\beta) && = \sin^2(\beta)  (1-\frac{\sin(q {\rm W})}{q {\rm W}}) +
9 \cos^2(\beta) e^{-\frac{{\rm W}^2}{{\rm l}^2}}
\\ \nonumber 
S_{p}^\prime(\beta) && = \cos^2(\beta)  (1-\frac{\sin(q {\rm W})}{q {\rm W}})
+ \frac{9}{2} \sin^2(\beta) e^{-\frac{{\rm W}^2}{{\rm l}^2}}
\end{eqnarray}
Similar expressions as in Eq. \ref{eq:ratePiez}  
but with  terms $G_{_{{\rm TA1}}}(x)$ and  $G_{_{{\rm TA2}}}(x)$ 
associated with ${\rm TA1}$ and ${\rm TA2}$ modes  
are obtained for  qubits interacting with
transverse acoustic phonons  
\begin{eqnarray}
\label{eq:gfun2}
G_{_{{\rm  TA1}}}(x) && = 
\frac{3}{2 x^2}-\frac{\sqrt{\pi } e^{-x} 
(2 x + 3)\text{Erfi}\left(\sqrt{x}\right)}{4 x^{5/2}} \\
\nonumber
G_{_{{\rm  TA2}}}(x) && = 
\frac{ \sqrt{\pi } e^{-x} 
(x (x (2 x + 5)+12)+15) \text{Erfi}\left(\sqrt{x}\right)}{6 x^{7/2}}-
\frac{x (x+2)+15}{3 x^3}
\end{eqnarray}
As expected $1/\tau_{_{\rm  Piez}}^{r(p)} = 0$ when ${\rm l_h = l_e}$
due to piezoelectric coupling being a polar mechanism.

%%%%%%%%%%%%%%%%%%
\begin{figure}[ht]
\includegraphics[width=11.4cm]{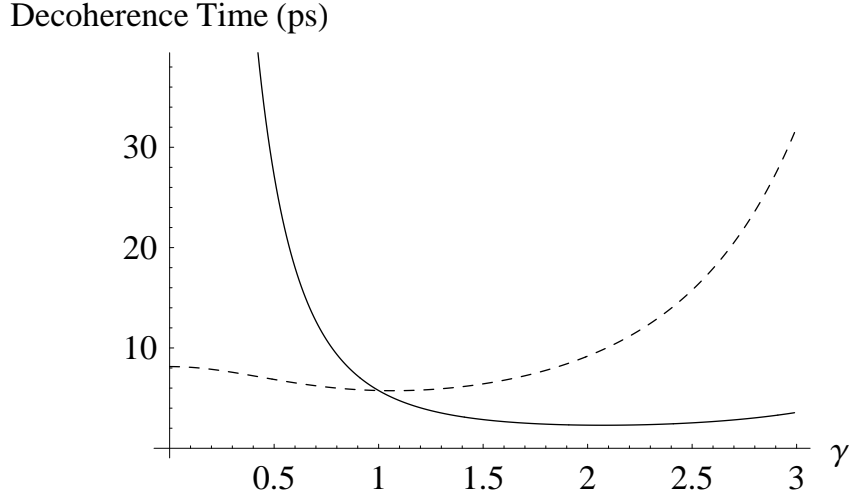}
\caption{Relaxation time, $\tau_{_{\rm  Piez}}^r$ (dashed)
and dephasing time $\tau_{_{\rm  Piez}}^p$ (full) as functions of 
$\gamma$ at $W$ = 5 nm, ${\rm l_e  = l_{ze}}$ = 2 nm, 
$r = {\rm l_h/l_e}$ = 5  and $T$ = 10 K.}
\label{fig:Piezgam}
\end{figure}
%%%%%%%%%%%%%%%%%%

%%%%%%%%%%%%%%%%%%
\begin{figure}[ht]
\includegraphics[width=11.4cm]{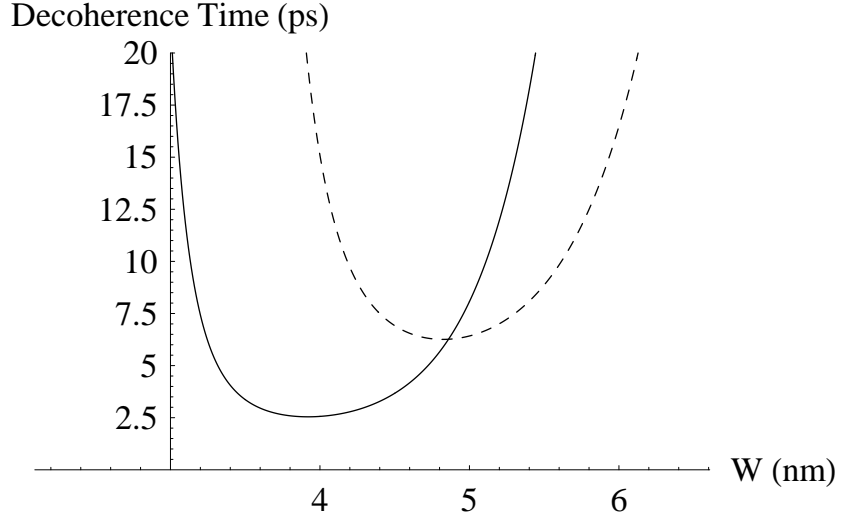}
\caption{Relaxation time $\tau_{_{\rm  Piez}}^r$
as function of interdot distance $W$ at $\gamma$ = 1.5 (dashed) and
$\gamma$ = 0.1 (full) at  ${\rm l_e  = l_{ze}}$ = 2 nm, 
$r$ = 5  and $T$ = 10 K.}
\label{fig:PiezW}
\end{figure}
%%%%%%%%%%%%%%%%%

Fig ~\ref{fig:Piezgam} shows that dephasing process due to 
 qubits interacting with phonons via deformation potential is the dominant
mechanism of decoherence for larger values of $\gamma$ ($\ge$ 2.5)
while the relaxation process dominates at small values of $\gamma$. 
As expected, these features are similar to those obtained in the
case of phonon coupling via deformation potential. 
These results as well as those obtained for the deformation potential
case highlight the critical role of 
$\gamma$ and  quantum dot parameters
in influencing the decoherence properties
of excitonic  qubit systems. Therefore the quality
factor $Q$ \cite{hon} which determines the number of charge oscillations that can be 
resolved within the decoherence time can be selected using  the quantum
dot system configuration based on 
parameters ${\rm l}$, ${\rm l_{z}}$, $r$, ${\rm W}$ and $T$. It appears that phonon 
assisted decoherence can be suppressed
by careful choice  of system parameters leading to higher fidelity of logic
gate operation. In the next section we consider another source of decoherence
involving one-phonon assisted F\"orster transfer process.

\section{\label{ForsterM}  Decoherence associated with 
 one-phonon assisted F\"orster transfer process}

The  decoherence rate, $\frac{1}{\Upsilon_{_{{\rm X},\lambda}}}$ 
 associated with phonon assisted F\"orster 
relaxation process is obtained using Eq. \ref{eq:rate} where  
$\hat H_{{\rm int}}$ is evaluated via
\begin{eqnarray}
\label{eq:Trmatrix}
{| \langle f| \hat H_{\rm F}^{\rm ph}|i \rangle|}^2 = \frac{1}{4}
|\sum_{i=1}^{4} T_{_{r,{\rm X}}}^i|^2
\end{eqnarray}
where  ${\rm X = DP \; \; or \; \; Piez}$
 and $T_{_{r,{\rm X}}}^i$ are the 
 transition amplitudes associated with
qubit relaxation. $T_{_{r,{\rm X}}}^1$
and $T_{_{r,{\rm X}}}^2$ for example  
are  obtained using Eqs. \ref{eq:pathrelax1} and \ref{eq:pathrelax3} 
\begin{eqnarray}
\label{eq:Trmatrix2}
T_{_{r,{\rm X}}}^1 && = \mp \frac{1}{\hbar \omega_{{\bf q}\; \lambda}} 
\bra{\chi_{\rm as}; n_{{\bf q}\; \lambda} \pm 1} 
\hat H_{F} \ket{\chi_{\rm as}; n_{{\bf q}\;\lambda}\pm 1} \;
 \bra{\chi_{\rm as}; n_{{\bf q} \; \lambda} \pm 1}
\hat H_{\rm ex-qb}^{\rm X}
 \ket{\chi_{\rm s}; n_{{\bf q}\; \lambda} } \\ 
T_{_{r,{\rm X}}}^4 && = \pm \frac{1}{\hbar \omega_{{\bf q}\; \lambda}} 
\bra{\chi_{\rm as}; n_{{\bf q}\; \lambda} \pm 1} 
\hat H_{\rm ex-qb}^{\rm X}
 \ket{\chi_{\rm s}; n_{{\bf q}\;\lambda}} \;
\bra{\chi_{\rm s}; n_{{\bf q} \; \lambda}}
\hat H_{F} \ket{\chi_{\rm s}; n_{{\bf q}\; \lambda} } 
\end{eqnarray}

In the case of dephasing mechanism, we use
\begin{eqnarray}
\label{eq:Tpmatrix}
{| \langle f| \hat H_{\rm F}^{\rm ph}|i \rangle|}^2 = \frac{1}{8} 
\left [|\sum_{i=1}^{4} |T_{_{p,{\rm s}}}^i|^2 - 
|\sum_{i=1}^{4} |T_{_{p,{\rm as}}}^i|^2 \right ]
\end{eqnarray}
where the  transition amplitudes $T_{_{p,{\rm s}}}^i$ are evaluated using
 Eqs. \ref{eq:pathphase1} - \ref{eq:pathphase4}. The
transition amplitudes $T_{_{p,{\rm as}}}^i$ associated with 
transitions between the $\chi_{\rm as}$ is obtained analogous to 
Eq. \ref{eq:Trmatrix}.

By neglecting the interference between terms in  Eqs. \ref{eq:Trmatrix}, we obtain 
explicit terms  for the rates of relaxation and dephasing 
\begin{eqnarray}
\label{eq:rateForst}
1/\tau_{_{\rm DP,F}}^{r(p)} 
= \frac{(D_c-D_v)^2 q^3}{32 \sqrt{\pi} \hbar  (1 + \gamma^2)\rho \upsilon_{_{{\rm LA}}}^2}
\exp(-\frac{3}{4} q^2 {\rm l}^2) \frac{{\rm Erfi}(\Delta {\rm l})}
{\Delta {\rm l}} (N_{{\bf q}, \lambda} +1) S_{r(p)}^F(\beta) \\ 
\end{eqnarray}
where  $\gamma = \frac{\Delta \Omega}{2 F({\rm W})}$ and the functions $S_{r}^F(\beta)$
and  $S_{p}^F(\beta)$ are given by 
\begin{eqnarray}
\label{eq:HexpF}
S_{r}^F(\beta) && = 
\sin^4(\beta)+ \frac{1}{2} \cos^2(\beta) (3 + \cos{2 \beta})
+ \sin(2 \beta)  e^{-\frac{{\rm W}^2}{{\rm l}^2}} +
\left [ \frac{1}{4}\sin^2 (2 \beta)- \sin^4(\beta) \right ]\frac{\sin(q {\rm W})}{q {\rm W}}
 \\ \nonumber 
S_{p}^F(\beta) && = 8 \sin^3(\beta)
\frac{\sin(q {\rm W}/2)}{q {\rm W}} e^{-\frac{{\rm W}^2}{2 {\rm l}^2}}
\end{eqnarray}
The  combined rates of  relaxation and dephasing for qubits coupled to
the   ${\rm LA}$, ${\rm TA1}$ and ${\rm TA2}$ modes  of the acoustic phonons
associated with the piezoelectric fields are  obtained as 
\begin{eqnarray}
\label{eq:ratePiezForst}
1/\tau_{_{\rm  Piez,F}}^{r(p)}
= \frac{\pi^{3/2} q e^2 e_{14}^2}{16 \epsilon^2 (1 + \gamma^2)
 \hbar \rho \upsilon_{_{{\rm LA}}}^2}\;  \exp(-q^2 {\rm l}^2/2)
\left [1-\exp(-\frac{(r^2-1) q^2 {\rm l}^2}{4})\right]^2
 G_{_{{\rm LA}}}(\frac{q^2 {\rm l}^2 r^2}{2 \left(r^2+1\right)})
 (N_{{\bf q}, \lambda} +1)
 S_{r(p)}^{F^\prime}(\beta)
\end{eqnarray}
where the function $G_{_{{\rm F}}}(x)$ is obtained using  Eq. \ref{eq:gfun}
and  associated functions, $G_{_{{\rm TA1}}}(x)$ and  $G_{_{{\rm TA2}}}(x)$
(see Eq. \ref{eq:gfun2})
and the functions $S_{r}^{F^\prime}(\beta)$
and  $S_{p}^F(\beta)$ are given by 
\begin{eqnarray}
\label{eq:HexpFp}
S_{r}^{F^\prime}(\beta) && = g_1(\beta) + g_2(\beta)  e^{-\frac{{\rm W}^2}{l^2}} + 
g_3(\beta)\frac{\sin(q {\rm W})}{q {\rm W}}
 \\ \nonumber 
S_{p}^{F^\prime}(\beta) && =g_4(\beta) 
\frac{\sin(q {\rm W}/2)}{q {\rm W}} e^{-\frac{{\rm W}^2}{2 l^2}}
\end{eqnarray}
where $g_i(x), i = 1, \ldots ,4$ can be obtained using  Eqs. \ref{eq:anfactorLA},
\ref{eq:anfactorTA1} and \ref{eq:anfactorTA2}.

%%%%%%%%%%%%%%%%%%
\begin{figure}[ht]
\includegraphics[width=11.4cm]{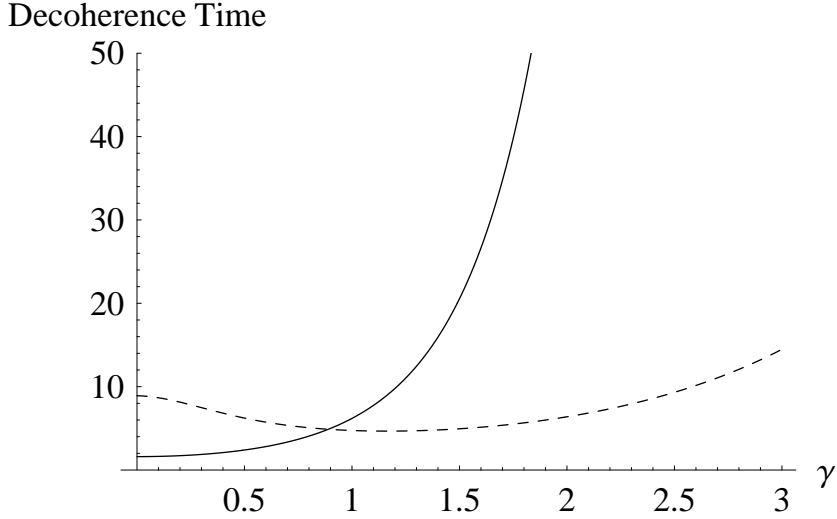}
\caption{Comparison of relaxation time $\tau_{_{\rm DP,F}}^r$ (dashed) and
dephasing time ${\tau_{_{\rm DP,F}}^{p}}$ (full) due to deformation potential coupling
as function of $\gamma$ at $W$ = 4 nm (full), ${\rm l}$ = 2 nm, ${\rm l_z}$ = 1 nm and
$T$ = 10 K.}
\label{fig:FosDP1}
\end{figure}
%%%%%%%%%%%%%%%%%%

%%%%%%%%%%%%%%%%%%
\begin{figure}[ht]
\includegraphics[width=11.4cm]{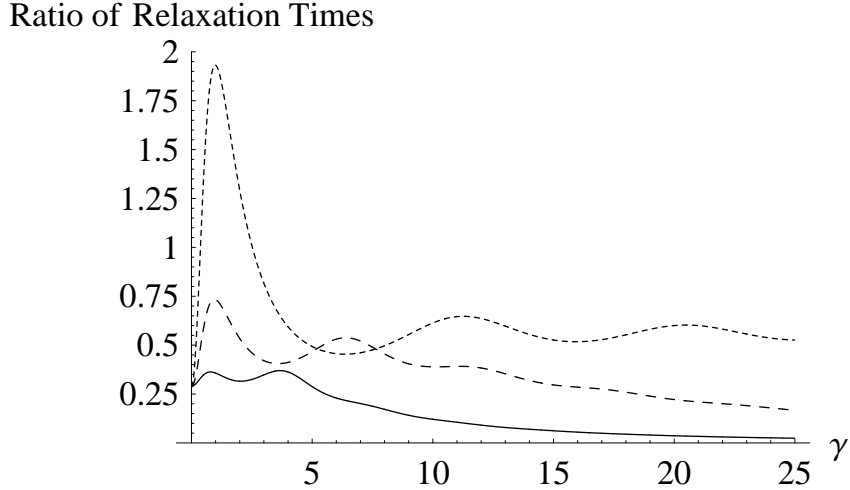}
\caption{Ratio of relaxation times $\frac{\tau_{_{\rm DP}}^r}{\tau_{_{\rm DP,F}}^{r}}$
as function of $\gamma$ at $W$ = 4 nm (full), 5 nm (dashed), 6.5 nm (dotted)
and ${\rm l}$ = 2 nm, ${\rm l_z}$ = 1 nm and
$T$ = 10 K.}
\label{fig:FosDP2}
\end{figure}
%%%%%%%%%%%%%%%%%%

%%%%%%%%%%%%%%%%%%
\begin{figure}[ht]
\includegraphics[width=11.4cm]{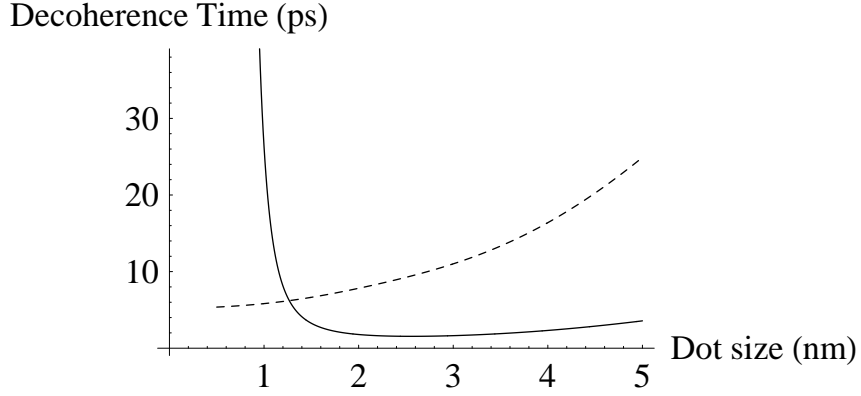}
\caption{Relaxation time, $\tau_{_{\rm DP,F}}^r$ (dashed)
and dephasing time $\tau_{_{\rm DP,F}}^p$ (full) as functions of 
dot size ${\rm l}$ at $\gamma$ = 0.25,  $W$ = 4 nm, ${\rm l_z}$ = 1 nm 
and $T$ = 10 K.}
\label{fig:FosDP3}
\end{figure}
%%%%%%%%%%%%%%%%%%

%%%%%%%%%%%%%%%%%%
\begin{figure}[ht]
\includegraphics[width=11.4cm]{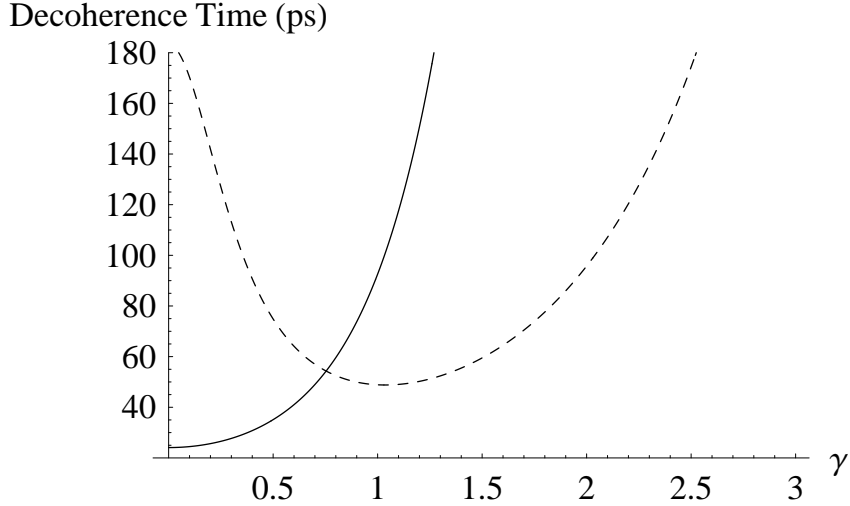}
\caption{Comparison of relaxation time $\tau_{_{\rm Piez,F}}^r$ (dashed) and
dephasing time $\tau_{_{\rm Piez,F}}^p$ (full) due to piezoelectric
field as function of $\gamma$ at ${\rm W}$ = 5 nm (full), 
${\rm l}$ = 2 nm, $r$ = 5 nm and
$T$ = 10 K.}
\label{fig:FosPiez1}
\end{figure}
%%%%%%%%%%%%%%%%%%

%%%%%%%%%%%%%%%%%%
\begin{figure}[ht]
\includegraphics[width=11.4cm]{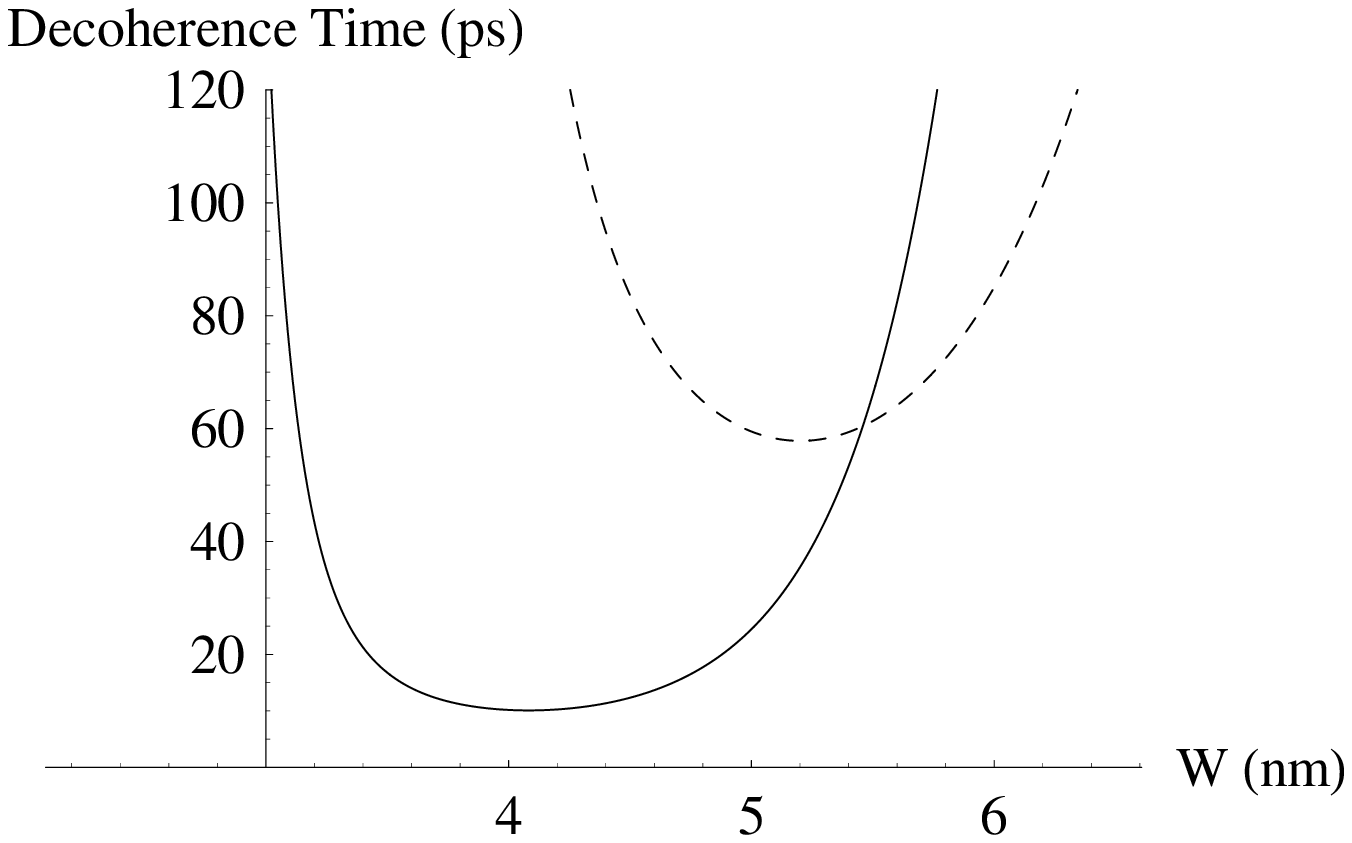}
\caption{Relaxation time $\tau_{_{\rm Piez,F}}^r$ 
 as function of interdot distance $W$ at $\gamma$ = 1.5 (dashed) and
$\gamma$ = 0.1 (full) at  ${\rm l}$ = 2 nm, $r$ = 5 nm  and
$T$ = 10 K.}
\label{fig:FosPiez2}
\end{figure}
%%%%%%%%%%%%%%%%%%

Fig.~\ref{fig:FosDP1} shows that pure dephasing dominates 
at low values of $\gamma$ in a  one-phonon assisted F\"orster transfer process
while the relaxation process is the 
 preferred mode  of decoherence at higher values of  $\gamma$. 
 These features are in reverse of 
those obtained in Fig.~\ref{fig:ratesDP} where the phonon mediated process
takes place in the vicinity of individual quantum dots.  
At very small values of $\gamma$, the decoherence times reach 
values of the order of several picoseconds showing the
significant loss of coherence which  occurs during 
  charge carrier oscillations between qubit states.
Fig.~\ref{fig:FosDP2} shows that relaxation via phonon-assisted
 F\"orster  transfer process is a more likely  route for decoherence
than  the direct mode of relaxation  at large interdot 
separations and small $\gamma$ values. The notable oscillations
in relaxation times is possibly  due to matching
of the interdot separation with multiples of the phonon wavevector
as  $\gamma$ is increased. In   Fig.~\ref{fig:FosDP3} we note 
the  reverse effects on the two types
of decoherence mechanisms due to an increase in size parameter ${\rm l}$

Fig.~\ref{fig:FosPiez1}  shows that the  one-phonon assisted 
F\"orster transfer process associated with 
piezoelectric fields proceeds in the same manner as in the model
involving deformation potential. Fig.~\ref{fig:FosPiez2}
highlights the range of ${\rm W}$ for which decoherence times
can match typical logic gate times of around 10 - 50 ps. In the
presence of an  external electric field,  
energy  for both electrons and holes is reduced with larger dots
experiencing bigger shifts.  We expect  changes to the order of
at most a magnitude to the results obtained here.

\section{\label{Concurr}  Berry phase and Concurrence of the composite 
exciton system}

We evaluate  two important quantities, namely
the Berry phase and concurrence of the composite 
exciton system in quantum dots. 
We consider a simplified model in which 
the  F\"orster interaction is switched off after  $t$= 0 and 
assume that there is no correlation between the phonon bath 
interacting with the correlated electron-hole
 pairs at different quantum dots. The  Hamiltonian in
Eq. \ref{eq:envqubit} can be solved \cite{arvind} 
to produce an explicit expression for the 
reduced density matrix   of the two qubits system at a later time $t$.
 The overall initial density matrix ($\rho_{T}(0)$) of the system of qubits
and phonon reservoir, $\rho_{ph}(0)$ is obtained as 
\begin{equation}
\label{eq:densT}
\rho_{T}(0) = \rho_{ex}(0)^{L} \otimes \rho_{ex}(0)^{R} 
\otimes \rho _{ph}^{L}(0) \otimes \rho_{ph}^{R}(0) \; .
\end{equation}
where $\rho_{ex}(0)^L$ and $\rho_{ex}(0)^R$ are the 
initial density matrix states of the two-state qubit system 
(see Eq. \ref{eq:qstates}).  $\rho _{ph}^{L}(0)$ and
$\rho _{ph}^{R}(0)$  are density matrices  associated with
the phonon reservoir in the quantum dots at $\ket{\bf R_{\rm a}}$ and  
$\ket{\bf R_{\rm b}}$,  respectively. At $t=0$, each phonon reservoir
is assumed to be thermalized at temperature $T$
\begin{equation}
\label{eq:phres}
\rho _{ph}^{I}(0)=\prod\limits_{\bf q}
( 1-e^{-\frac{\omega_{\bf q}}{k_B T}} )
e^{- \frac{ \omega_{\bf q}}{ k_B T} b_{\bf q}^{\dagger} \; b_{\bf q}}  \;   .
\end{equation}
where $I = L,R$. 

We consider an initial state  given by $\ket{\chi_{\rm s}}
= \cos(\beta(0)/2) \ket{\bf L} \; + 
\sin(\beta(0)/2)\ket{\bf R}$ (see  Eq. \ref{eq:statesEigen})
which depends on the interdot
F\"orster interaction amplitude $F$ at $t$ = 0 through the angle
$\beta(0)$.  We will show later that the evolution of $\beta$ with
time is dependent on the dynamics of interaction between the 
qubit states and phonon bath.
In the case of identical qubits, the state 
$ \rho_T(t) = \ket{\chi_{\rm s}} \bra{\chi_{\rm s}}$ evolves
 according to 
\begin{gather}
\label{eq:matconc}
\rho _T(t) =\left(
\begin{array}{cccc}
0 & 0 & 0 & 0 \\
0 & \cos^2(\beta /2) & \frac{1}{2}\sin(\beta(0))\, 
e^{ {\rm i} \Delta \Omega t} Z(t) & 0 \\
0 & \frac{1}{2}\sin(\beta(0))\, e^{-{\rm i} \Delta \Omega t}
Z(t) &\sin^2(\frac{\beta(0)}{2}) & 0 \\
0 & 0 & 0 & 0
\end{array}
\right) \; .
\end{gather}
where $\Delta \Omega$  is the difference in  exciton creation 
energy between the two quantum dots and  
$Z(t) = e^{-4\left[ G_{L}(t) +G_{R}(t) \right] }$ 
provides a  measure of  the decay of the off-diagonal matrix
elements when each qubit is isolated from the other, but exposed
to its own reservoir of phonon bath. The function
$G(t) = \exp( -2 \int \frac{d \omega}{\hbar \omega^2} 
J(\omega) \coth(\hbar \omega/2 k_B T)
\sin^2(\omega t/2))$ where $J(\omega)$ is 
the spectral density function which yields 
information about the interaction of the quantum dot with
the phonons \cite{Leggett}
\begin{eqnarray}
\label{eq:specdenDP}
J_D(\omega) && = \sum _{\bf{q}} \Xi_D(q_\parallel,q_z)^2 
\delta(\omega -\omega_{\bf{q}})  \quad \quad {\rm Deformation \; \; Potential}
\\ \label{eq:specdenPiez}
J_P(\omega,\lambda) && =  \sum _{\bf{q}} \Sigma_P(q_\parallel,q_z,\lambda)^2 
\delta(\omega -\omega_{\bf{q}})   \quad \quad {\rm Piezolectric \; \; Interaction}
\end{eqnarray}
where $\Xi_D(q_\parallel,q_z)$ and $\Sigma_P(q_\parallel,q_z,{\rm LA})$ are given in  
Eqs. \ref{eq:funcDP} and \ref{eq:funcPi} respectively. 
At small $\omega $, $J(\omega)\sim \omega ^{k}$ where  the exponent
$k$ determines between the cases of ohmic ($k=1$), sub- ($k < 1$) and superohmic
($k > 1$) couplings \cite{krumm}. 
The explicit expressions for $J_D(\omega)$ and  $J_P(\omega,{\rm LA})$
 are determined  using Eqs. \ref{eq:PmatrixDPr2} 
and \ref{eq:PmatrixPi2} to be in superohmic forms where 
$k=3$ and $k=5$  respectively for $J_D(\omega)$ and  $J_P(\omega,{\rm LA})$.

The Berry phase  of a mixed state is defined as
\begin{equation}
\label{eq:gp}
\eta = {\rm arg} \left[ \sum_{j} \sqrt{E_{j}(0) E_{j}(T)} 
\langle \chi_j(0)|\chi_j(T) \rangle
e^{-\int_0^T dt \langle \chi_j(t) | \dot{\chi_j}(t)\rangle}  \right ]
\end{equation}
where  the system frequency $\Delta \Omega$  determines the quasi-cyclic path
 of the geometric phase with time $t$  varying from 0 to $T = \frac{2 \pi}{\Delta \Omega}$.
$E_{j}(T)$ and $\chi_j(T)$ refer to the eigenvalues and eigenvectors, respectively of
the  density matrix  in  Eq. \ref{eq:matconc} 
\begin{eqnarray}
\label{eq:mateigen}
E_{\pm}(t) && = \frac{1}{2} \pm \frac{1}{2} 
\left( \cos^2(\beta) +  \exp(-2 G(t)) \sin^2(\beta) \right )^2
\\ \nonumber
\ket{\chi_{+}(t)} && = \cos(\frac{\beta(t)}{2}) \ket{\bf R} \; + 
\sin(\frac{\beta(t)}{2})\ket{\bf L} 
\end{eqnarray}
where $\tan(\frac{\beta(t)}{2}) = \exp(-G(t)) \cot(\frac{\beta(0)}{2})$. 
and  the evolution of the phase is determined only by  
$\ket{\chi_{+}(t)}$ as $E_{-}(0)$ = 0. At $t=$ 0, $G(t) =$ 0, $Z(t) =$ 1
and $\ket{\chi_{+}(t)} \rightarrow \ket{\chi_{\rm s}(0)}$. Substituting 
Eq. \ref{eq:mateigen}  into  Eq. \ref{eq:gp} we evaluate 
$\eta = \eta_0 + \eta_c$ where $\eta_0 = \pi (1 - \cos(\beta))$
is the well known result for a unitary evolution at $G(t)$ = 0.
The unitary phase correction term  $\eta_c$ which appears when $G(t) \neq 0$ 
is evaluated using series expansion of $G(t)$. 

%%%%%%%%%%%%%%%%%%
\begin{figure}[ht]
\includegraphics[width=11.4cm]{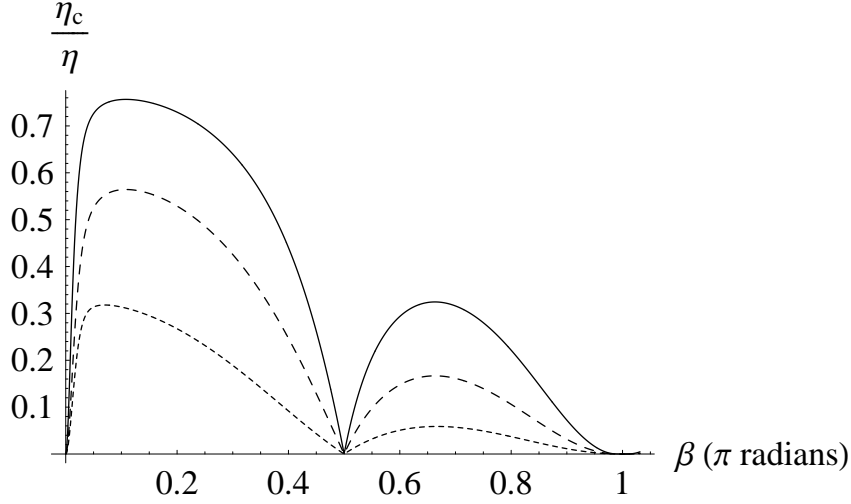}
\caption{Berry phase correction fraction $\frac{\eta_c}{\eta}$
 as function of polar  angle $\beta$ due to decoherence associated
 with deformation potential at $W$ = 5 nm, 
${\rm l}$ = 1 nm and $T$ = 120 K (full), 50 K (dashed) and 
 $T$ = 0 K (dotted)} 
\label{fig:gp1}
\end{figure}
%%%%%%%%%%%%%%%%%%

%%%%%%%%%%%%%%%%%%
\begin{figure}[ht]
\includegraphics[width=11.4cm]{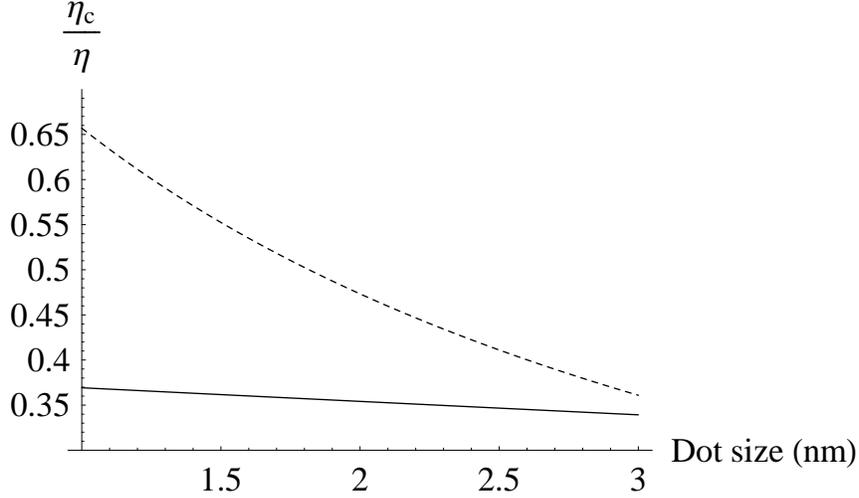}
\caption{Berry phase correction fraction $\frac{\eta_c}{\eta}$
 as function of the dot size due to decoherence associated
 with deformation potential (dashed)
 and piezoelectric coupling (full) at $W$ = 5 nm, 
$\beta = \frac{\pi}{4}$ and $T$ = 100 K.}
\label{fig:gp2}
\end{figure}
%%%%%%%%%%%%%%%%%%

%%%%%%%%%%%%%%%%%%
\begin{figure}[ht]
\includegraphics[width=11.4cm]{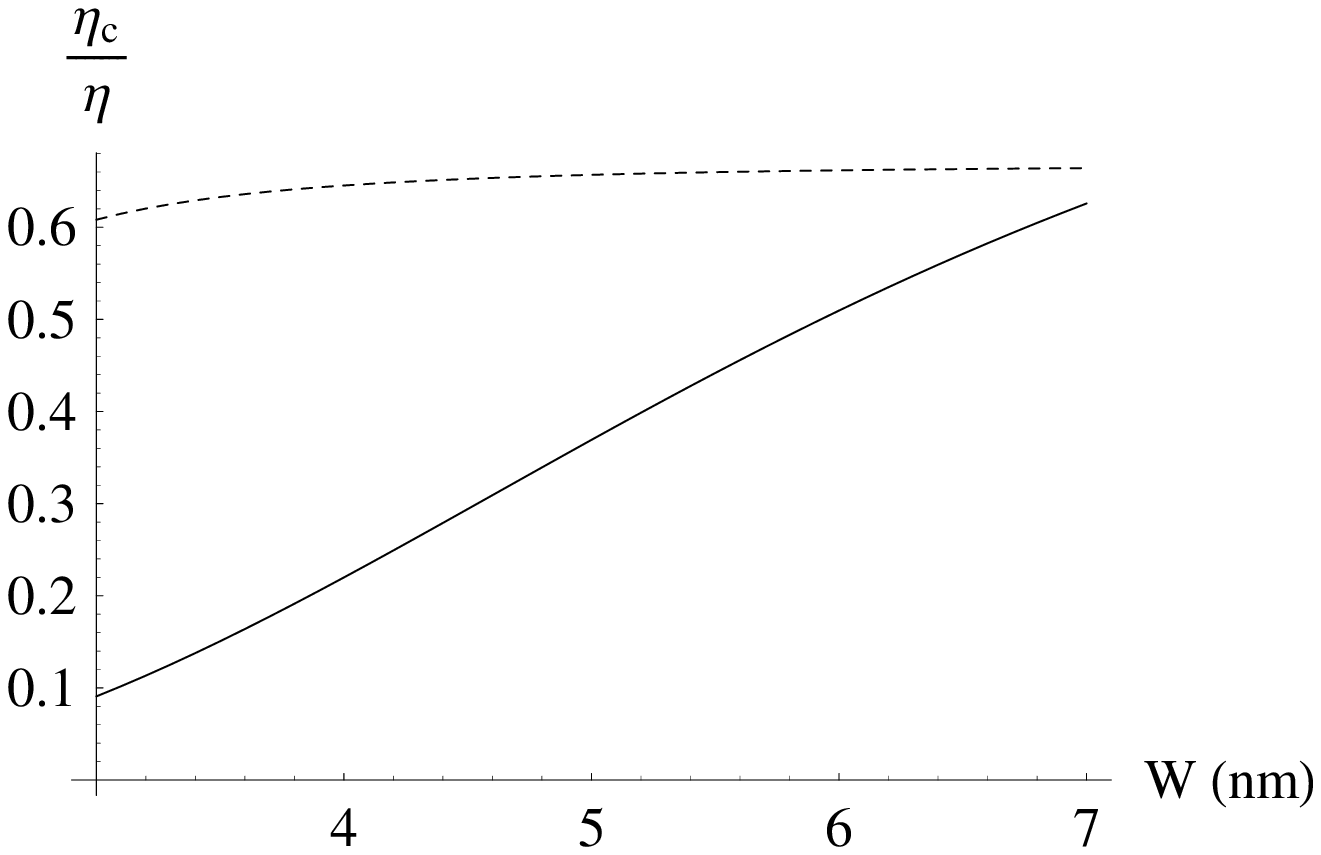}
\caption{Berry phase correction fraction $\frac{\eta_c}{\eta}$
 as function of interdot distance  ${\rm W}$ 
 due to decoherence associated  with deformation potential(full)
 and piezoelectric coupling (dashed) at $W$ = 5 nm, 
$\beta = \frac{\pi}{4}$ and $T$ = 100 K.}
\label{fig:gp3}
\end{figure}
%%%%%%%%%%%%%%%%%%

 Fig.~\ref{fig:gp1}  shows the   increase of 
Berry phase correction fraction $\frac{\eta_c}{\eta}$
with temperature in  GaAs/AlGaAs quantum dots 
coupled to  phonon baths via deformation  potential. 
The geometric phase of the qubit state undergoes a greater degree
 of nonunitary evolution within a superohmic 
environment at larger temperatures. At $T=$ = 0 K, zero point 
fluctuations associated with  a 
noisy environment contributes to the non-zero correction in
Berry phase. At $\beta = n \pi$ where $n$ is an integer,
 $T = \frac{2 \pi}{\Delta \Omega}$ vanishes so that the
system stagnates and  $\frac{\eta_c}{\eta} = $ 0 at the
nodal points. At  $\beta = n \frac{\pi}{2}$,  $T$ becomes
very large and that the system  never completes its quasi-cyclic path
and there is negligible phase correction. At other values of 
 $\beta$, the departure from unitary behavior is
clearly  dependent on the spectral density function  $G(t)$
as the Bloch sphere is altered 
according to  $G(t)$. There is  decreased effect of  decoherence effects on $\eta$
as the phase angle sweeps the second  quarter 
to complete the first half of cyclic evolution.

 Fig.~\ref{fig:gp2} shows the gradual decrease of  Berry phase correction 
and hence progress towards a more unitary evolution at larger quantum dot sizes
for both types of phonon couplings. The difference between the two curves
can be attributed to microscopic properties of the spectral 
density functions $J_D(\omega)$ and  $J_P(\omega)$. In both cases, the
density function  rises rapidly  with $\omega$ and reach peak values at an 
optimum $\omega_m$ before starting to decrease at even higher values of $\omega$.
Fig.~\ref{fig:gp3} shows the increase of  Berry phase correction with
interdot distance  ${\rm W}$ which gives a measure of the degree of entanglement
at $t=$ 0. It is to be noted that ${\rm W}$ provides a 
direct link between concurrence  (details below)
and correction to the  Berry phase. At large ${\rm W}$, 
the concurrence of  the qubit states at $t=$ 0 is weak
 and the Berry phase is less immune
to decoherence effects  and undergoes a higher 
degree of non-unitary evolution at $t \ge$ 0.

The evolution of the entanglement between the subsystems 
 $\rho_{ex}(0)^L$ and $\rho_{ex}(0)^R$ is determined by a
quantity termed concurrence, $C(t)$ which is  related to the 
entanglement of formation \cite{woott}. For a pure or
 mixed state, $\rho_T$, of two qubits, the spin-flipped state is defined as
\begin{gather}
\widetilde{\rho}_T = \left( \sigma _y\otimes \sigma _y\right) \rho_T
^{*}\left( \sigma _y \otimes \sigma _y\right) \; ,
\end{gather}
where $\sigma _y$ belongs to the set of Pauli matrices.
The concurrence is given by  \cite{woott}
\begin{gather}
\label{eq:concw}
C\left( \rho_T(t) \right) = \max \big\{0,\;2\max\limits_i \lambda _i
 - \sum\limits_{j=1}^4 \lambda_j\big\} \; .
\end{gather}
where $\lambda _i$s are the square roots of the eigenvalues of 
the non-Hermitian matrix ${\rho}_T \widetilde{\rho}_T$ with 
each value of $\lambda _i$ being a non-negative real number.

The  eigenvalues of the matrix product ${\rho}_T \widetilde{\rho}_T$
for the density matrix  Eq. \ref{eq:matconc} are
\begin{equation}
\label{eq:conceigen}
\lambda_{1,2}=\frac{1}{2} \sin(\beta) (Z(t) \pm  1),
\end{equation}
and $\lambda_{3,4}=0$. Hence the concurrence is obtained as
\begin{equation}
\label{eq:concfi}
C(t) =   \frac{1}{2} \sin(\beta) e^{-4\left[ G_{L}(t) +G_{R}(t) \right]}
\end{equation}

%%%%%%%%%%%%%%%%%%
\begin{figure}[ht]
\includegraphics[width=11.4cm]{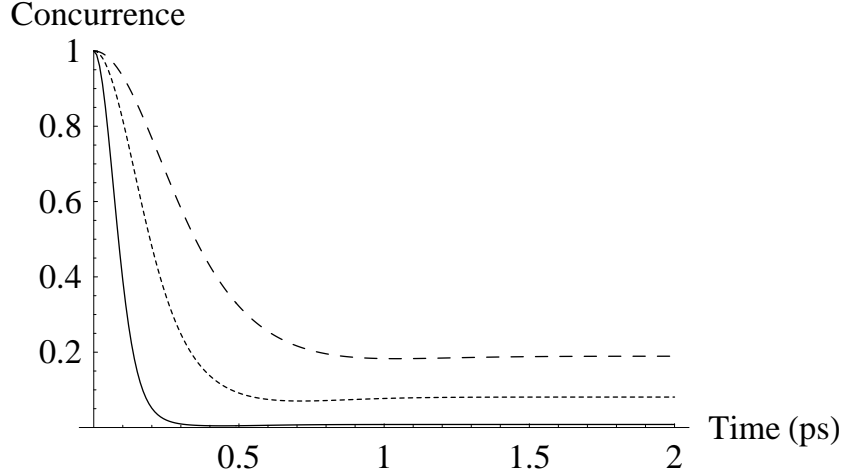}
\caption{Concurrence decay in
qubits coupled to uncorrelated phonon baths  
via deformation potential coupling  at ${\rm l}$ 
= 2 nm (dashed), 1 nm (full), 1.5 nm (dotted)
and $T$ = 10 K, with the prefactor
 $\frac{1}{2} \sin(\beta)$ suppressed.}
\label{fig:conc1}
\end{figure}
%%%%%%%%%%%%%%%%%%

%%%%%%%%%%%%%%%%%%
\begin{figure}[ht]
\includegraphics[width=11.4cm]{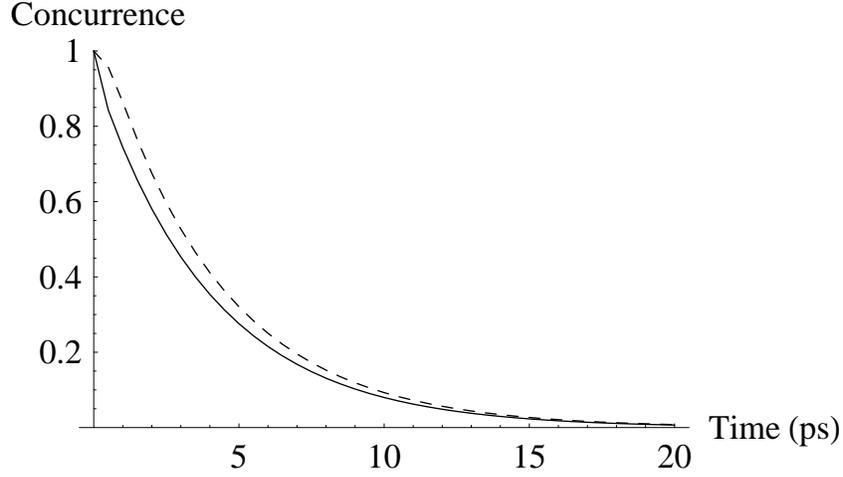}
\caption{Concurrence decay in 
qubits coupled to uncorrelated phonon baths  
via piezoelectric coupling  at ${\rm l}$ 
= 3 nm (dashed), 0.5 nm (full) and $T$ = 10 K,
 with the prefactor 
$\frac{1}{2} \sin(\beta)$ suppressed.}
\label{fig:conc2}
\end{figure}
%%%%%%%%%%%%%%%%%%

 Figs.~\ref{fig:conc1} and ~\ref{fig:conc2} show  differences
in  half times of  concurrence decay  in GaAs/AlGaAs quantum dots which are
 isolated from each other but each exposed to their
 own phonon baths via deformation  potential and piezoelectric coupling.  
The half time during which half of the initial entanglement at $t=$ 0 
is lost is less for qubits interacting with phonons via  deformation potential compared
the qubit-phonon system incorporating piezoelectric coupling. These differences
can be attributed  to  functions $G_D(t)$ and  $G_P(t)$.
The function $G_D(t)$ associated with  the spectral 
density functions $J_D(\omega)$ (deformation  potential) rises with time and reaches
a constant value whereas $G_P(t)$ retains its monotonic increase
with time.   Both figures show an increase of half time with quantum dot 
size ${\rm l}$ which is due to  the shorter time needed
for the phonons to travel the length of the quantum dot and
decrease of  spectral density function $J(\omega)$ as the dot size is
increased. It is  to be noted that  in our model we have considered 
lattice vibrations at the two qubit locations to be  uncorrelated so
that the bath modes are independent  unlike the model
incorporating a common environment used by  Braun \cite{braun}.

\section{\label{Conclu}  Conclusions }

We have studied several processes that contribute to the
decoherence  of excitonic qubits in  quantum dot systems coupled by the 
F\"orster-type transfer process. We show the significant loss of
 coherence  that  occurs through charge carrier oscillations 
between the different  qubit states by using a model of
one-phonon assisted F\"orster-type transfer process. 
It is  shown that increasing a tuning factor $\gamma$ has opposite effects
on the relaxation and dephasing times for excitonic qubits interacting with 
acoustic phonons  via both deformation potential and piezoelectric coupling.
Our results show that  entanglement of qubits which are isolated 
from each other last for comparatively long times 
when interacting with uncorrelated phonons baths  via deformation potential
compared to the case of piezoelectric coupling which become 
completely disentangled within 20 ps. 
The results obtained here show that phonon mediated interactions 
 exert a sizable influence on the non-unitary
evolution  of Berry phase in quantum dot systems.
Lastly, our results  emphasise  the strong correlation between the 
Berry phase correction fraction which quantify departure from unitary
 evolution and half times of concurrence decay  in GaAs/AlGaAs quantum dots.

\end{document}